\documentclass[reprint,NumberedRefs]{JASA}

\usepackage{amsmath,amssymb,amsfonts}
\usepackage{siunitx}
\usepackage{multirow}
\usepackage{tabularx}
\usepackage{mathtools}
\usepackage{wasysym}
\usepackage{filecontents}

\begin{document}

\title{Preprint: Global, and Local Optimization Beamforming for Broadband Sources}

\author{Armin Goudarzi}
\email{armin.goudarzi@dlr.de}

\affiliation{German Aerospace Center (DLR), Germany}

\begin{abstract}
This paper presents an alternative energy function for Global Optimization (GO) beamforming, tailored to acoustic broadband sources. Given, that properties such as the source location, multipole rotation, or flow conditions are parameterized over the frequency, a CSM-fitting can be performed for all frequencies at once. A numerical analysis shows that the nonlinear energy function for the standard GO problem is equivalent to the source's Point Spread Function (PSF) and contains local minima at the grating- and side lobes' locations. The energy function is improved with the proposed broadband energy, as it averages the PSF. Further, it simplifies the process of identifying sources and reconstructing their spectra from the results. The paper shows that the method is superior on synthetic monopoles compared to standard GO and CLEAN-SC. For real-world data the results of the proposed method and CLEAN-SC are similar, and outperform standard GO. The main difference is that source assumption violations cause noisy maps for CLEAN-SC and cause wrong spectral estimations of the proposed method. By using reasonable initial values, the GO problem reduces to a Local Optimization problem with similar results. Further, the proposed method is able to identify synthetic multipoles with different pole amplitudes and unknown pole rotations. 
\end{abstract}
\maketitle

\section{Introduction}
Multiple noise-generating phenomena and mechanisms exist. For the localization and estimation of the sound power of complex source geometries, beamforming (referring to any covariance matrix-based imaging method) is well-established~\citep{MerinoMartinez2019a}. It was shown that an unknown, compactly supported source power function is uniquely determined by idealized, noise-free correlation measurements in a bounded measurement domain~\citep{Hohage2020}. However, it is well known, that correlated sources are not uniquely determined from distant measurements of acoustic waves since there exist so-called non-radiating sources. Additionally, most methods are grid-based. This means, that in the area where sources are expected, predefined focus points are generated, on which the beamforming method is then evaluated. Due to the increasing computational power, many algorithms are now evaluated on 3D problems, where the number of focus points is typically very large compared to the number of microphones so that even a solution for radiating sources is no longer unique.

Beamforming relies on several assumptions to obtain a unique source solution. The main assumptions generally include one or all of these: \replaced[R2C10]{spatially compact}{point-like} sources, monopole sources, incoherent sources, and independent sound radiation for each frequency, as well as undisturbed sound propagation through the medium, which are typically violated in real-world scenarios~\cite{Ahlefeld2010,Bahr2017,Martinez2020,Lehmann2022}.

\replaced{Naive}{Forward} methods such as Conventional Beamforming (CB) are widely popular due to their robustness and the knowledge of their limitations~\citep{MerinoMartinez2019a} and do not rely on uniqueness, as they estimate a source strength for each focus point independently. \added[R1C06]{Methods that adapt the choice of steering vectors based on the measurement such as robust adaptive beamforming~\hbox{\citep{Cox1987}} can improve the resolution but still suffer from the mismatch of source assumptions and the data itself~\hbox{\citep{Suzuki2011}}, especially in high noise situations such as closed wind tunnel measurements.} More sophisticated approaches exist, such as inverse methods~\citep{Suzuki2011,Zavala2011}, 'deconvolution' methods such as CLEAN-SC~\citep{Sijtsma2007}, and DAMAS~\citep{Brooks2006,Chardon2021b} where a sparse source distribution is reconstructed from beamforming maps to improve the spatial resolution. However, inverse methods and advanced deconvolution methods are often computationally expensive and still include assumptions about the source, due to the mismatch of independent equations and estimated variables. Compressed sensing~\citep{Edelmann2011,Xenaki2014} tries to relax this mismatch by assuming sparsity, however, sparsity is typically violated by the basis mismatch~\citep{Chi2011,Duval2017}. The incorporation of properties such as coherence and non-compactness imposes additional \replaced[R1C05,R2C07]{free variables}{unknowns}, which is typically relaxed with smoothing functions~\citep{Blacodon2004,Funke2012,Oertwig2022}.

Recently, gridless methods emerged\replaced[R1C02]{ that}{, which} have several advantages over grid-based methods. In these gridless methods, the focus points are no longer fixed in space. Thus, the inverse problem becomes nonlinear, but the number of unknown variables is reduced by multiple orders of magnitude, since fewer focus points are necessary to account for the sources, due to their sparseness compared to the domain. These methods include subspace beamforming~\citep{Sarradj2022}, machine learning based methods~\citep{Kujawski2022}, and Covariance Matrix Fitting (CMF) methods~\cite{Chardon2021a,Chardon2021b,Chardon2022a,Chardon2023}. 

Global Optimization (GO) is a gridless CMF method, employing a global optimizer such as Differential Evolution~\citep{Malgoezar2017,Hoff2022}. In this method, a synthetic CSM is directly fitted against an observed CSM based on a Mean Squared Error (MSE) energy function, that is minimized based on varying the estimated source positions and source strengths. While the method is straightforward, global optimizers are computationally expensive and not guaranteed to find the global minimum~\citep{Neumaier2004,Chardon2023}. Since CMF methods do not impose any limitations on the model, it is possible to include and optimize parameters that are often neglected in beamforming methods. These are source model assumptions, such as distributed sources~\citep{Dong2016,Blacodon2004}, the incorporation of wind tunnel flow effects, such as shear layer decorrelation~\citep{HaxterSpehr2012,Haxter2014,Sijtsma2014,ErnstSpehrBerkefeld2015,Ernst2020}, partially coherent source regions~\citep{Yardibi2010,MohanK2017,Leclere2017,Oertwig2022,Goudarzi_Bebec2022,Chardon2022c}, imposing real steering vectors through a measured or simulated flowing medium~\citep{MohanK2017,Lehmann2022}, the speed of sound~\cite{Ahlefeld2010,AhlefeldtKoop2010,Malgoezar2017}, and reflections~\citep{Lauterbach_etal2010}, correlated in-flight or closed wind tunnel shear layer noise~\citep{Ahlefeldt2021,HaxterSpehr2016,Haxter201785}, beamforming on multiple CSMs at different subsonic flow speeds, assuming self-similarity and a Mach power law~\citep{Ahlefeldt2013,Ahlefeldt2016,Goudarzi2022}, 
multipoles such as dipoles and quadrupoles, and source directivity~\citep{Funke2012,Oertwig2022,Goudarzi_Bebec2022}. Out of these properties, we will include multipoles in the source model as proof of concept for this paper. 

Quasi-stationary aeroacoustic sound sources are typically geometry-driven (e.g. leading, and trailing edges). Thus, the resulting source locations \added{and orientations} are constant over frequency~\citep{Goudarzi2021}. In this paper, we use this observation to introduce source objects, that have multiple properties, such as a location, and dipole rotation, which are constant over frequency. We introduce a modified broadband energy function to GO that allows the CSM-fitting of all frequencies at once, which makes use of the source and propagation parameters, that are shared for all frequencies. This paper shows, that the energy is driven by the Point Spread Function (PSF), which, depending on the problem, can lead to strong local minima. The proposed broadband energy averages the MSE over all frequencies which is equivalent to averaging the PSF and results in a much smoother energy function.

So far it is a drawback for beamforming methods \deleted[R1C02]{is} that the resulting Power Spectral Density (PSD) is obtained for each frequency $f$ independently. The resulting high-dimensional $\text{PSD}(f,x,...)$ is difficult to visualize and analyze over the three-dimensional space $x$, which is why it is typically spatially integrated to source spectra $\text{PSD}(f)$. Thus, a post-processing step is necessary to identify the source positions in the map, and the integration regions, so-called Regions of Interest (ROI). This is not trivial, and there exist a variety of methods to identify sources and extract their corresponding spectra from the beamforming results~\citep{Brooks1999,MerinoMartinez2019b,Goudarzi2021,Kujawski2022}. This paper discusses how spectra are extracted for CLEAN-SC and Global Optimization, as well as for the proposed method.

The parameterization of variables over the frequency (i.e. the source location and dipole rotation are constant over frequency. One might incorporate more advanced functions, that model physical properties here) increases the number of equations to unknown variables in the energy function, which makes the method a suitable candidate for the incorporation of the before-mentioned neglected source and propagation properties. Due to the computational complexity of the problem, it is of interest if one can transition from Global Optimization towards Local Optimization with reasonable initial conditions. This paper discusses this approach on real open wind tunnel data.

The paper is structured as follows. Sec.~\ref{sec:meth} introduces the source object, the source model, and the propagation operator. The CSM-fitting process and the corresponding GO energy function are presented. The energy function is then explored numerically for incoherent, \replaced[R2C10]{spatially compact}{point-like} monopoles, due to the nonlinearity of the inverse problem. A brief description of the source identification and spectrum generation problem based on beamforming results is provided. Sec.~\ref{sec:results} presents results for \replaced[R2C10]{spatially compact}{point-like}, synthetic monopole sources, real monopole-like sources in an open wind tunnel, and synthetic multipole sources. Sec.~\ref{sec:conclusion} summarizes the main findings.

\section{Methodology}\label{sec:meth}
This section introduces source objects, presents the sound propagation operator $\textbf{T}$, and the energy function $E$ for the CSM-fitting. Further, it presents the spectra generation problem from beamforming results and typical solution approaches.

\subsection{Source object}
A source object is a description of a single acoustic source $s$ with a collection of properties. These properties may vary from source object to source object and are all expressed by a finite number of numerical variables. For this paper, we will focus on \replaced[R2C10]{spatially compact}{point-like}, incoherent monopoles and dipoles, \added[R2C05]{but the incorporation of high order multipoles and distributed, coherent sources is possible as well}.

A monopole source object has the following properties: A fixed location $y$, and a PSD$(f)$. A dipole has the additional properties: The spatial dipole rotation angles $\theta, \varphi$, and the dipole PSD. Together, they can form a multipole source object with a single location, rotation, and separate monopole and dipole PSDs.

\subsection{Source description and propagation operator}
GO is a CSM fitting method (CFM). Thus, the free source model variables are optimized in such a way that the estimated CSM coincides with the measured CSM. For this process, a description of the source and the propagation is necessary. The acoustic propagation can be formulated as
\begin{equation}\label{eq:forward_operator}
    \textbf{T}q=c \quad,
\end{equation}
with $\textbf{T}$ being the propagation operator, $q$ being the (squared) source amplitudes, and $c$ being the vectorized CSM $\textbf{C}$. The propagation operator is derived from the Green's Matrix $\textbf{H}$ with
\begin{equation}
    \textbf{T}=\textbf{H}^* \odot \textbf{H}\quad,
\end{equation}
where $\odot$ is the column-wise Khatri-Rao product~\citep{Vanderveen1996} \added[R1C04]{and $*$ is the complex transpose}. The Green's matrix $\textbf{H}$ is given by 
\begin{equation}\label{eq:Greens_matrix}
    {H}_{mn} = h(x_m, y_n) \quad n=1, \dots, N \quad m=1, \dots, M \ .
\end{equation}
where $h$ describes the acoustic propagation from \replaced[R1C04]{a}{$N$} source locations $y$ to \replaced[R1C04]{a}{$M$} microphone locations $x$. For a free-field monopole, the propagation function $h$ is given by the Green's Function 
\begin{equation}\label{eq:monopole_green}
    h_\text{mono}(x, y) = \frac{\exp{(-jkd)}}{4\pi d} \ ,
\end{equation}
with the wavenumber $k=2\pi f/a$, $a$ is the speed of sound, $f$ is the frequency, and $d = |x-y|$ is the distance between the sources and sensors. For a dipole, the propagation function is given by
\begin{equation}\label{eq:dipole_green}
   h_\text{dip}(x_n, y) = \left({e}_\text{dip}\cdot {e}_n\right)\frac{\exp{\left(-jkd\right)}}{4\pi}\left(\frac{1}{d^2}+\frac{jk}{d}\right)
\end{equation}
with ${e}$ being the normalized direction vectors of the dipole and the microphone position $n$. The normalized vectors are given in spherical coordinates, which makes it easier to control the rotation and strength of the dipole independently with
\begin{equation}
    {e} = \left[ \sin(\theta)\cos(\varphi),\sin(\theta)\sin(\varphi),\cos(\theta)\right]^T \,.
\end{equation}
In traditional inverse methods~\citep{Suzuki2011} we find $q$ with
\begin{equation}
    q=\textbf{T}^{-1}c \quad,
\end{equation}
which requires typically a regularization due to the rank of $\textbf{T}$ (the propagation matrix is under-determined due to the exceeding number of focus points compared to the number of sensors, and the introduced regularization relaxes this problem by introducing spatial sparsity). In GO we need to find $q$ and $\textbf{T}$. \replaced[R209]{E.g., For spatially compact, incoherent monopoles the free variables}{For example the unknowns of point-like, incoherent monopoles} in $\textbf{T}$ are the source positions $y_n$, for dipoles there are the additional dipole orientation angles $\theta_n$, and $\varphi_n$, see eq.~\ref{eq:monopole_green} and eq.~\ref{eq:dipole_green}. Thus, there are four \replaced[R1C05,R2C07]{free variables}{unknowns} per monopole, and six \replaced[R1C05,R2C07]{free variables}{unknowns} for dipoles. Additionally, there are $F$ \replaced[R1C05,R2C07]{free variables}{unknowns} in $q$ for each multipole PSD, where $F$ is the number of employed frequencies.

\subsection{Energy function and optimization}\label{sec:error}

To optimize $q$ and $\textbf{T}$ we need an energy function $E$, also referred to as a cost function. Von den Hoff et al.~\citep{Hoff2022} suggest the Mean Square Error (MSE) between the measured CSM $c_{\text{meas}}$ and the estimated CSM $c_{\text{mod}}$, that is
\begin{equation}\label{eq:E_standard_GO}
    E(f) = \sum_{\hat{\imath}} \left( |c_{\text{mod},\hat{\imath}}(f)-c_{\text{meas},\hat{\imath}}(f)|^2 \right) \quad.
\end{equation}
Only elements $\hat{\imath}$ that correspond to the upper triangular CSM without diagonal should be evaluated to exclude uncorrelated self-noise\added[R2C11]{, so that the CSM in matrix form $C_{ij}$ is summed for $i>j$}. Note, that according to eq.~\ref{eq:forward_operator} $c_\text{mod}$ is the superposition of the uncorrelated source CSMs. 
Here, the energy function is evaluated for each frequency separately. We will reformulate the energy function so that it is evaluated for all frequencies at the same time. Since we will compare the energy for a varying number of frequencies and microphones, we replace the sum by an average $\langle\dots\rangle$ over all employed frequencies and upper CSM entries. The average subscript indicates which variable will be averaged. An additional normalization for each frequency is added to account for different source strengths at different frequencies. The new energy function then is
\begin{equation}\label{eq:cost_function}
    E = \left\langle \frac{ |c_{\text{mod},\hat{\imath}}(f_j)-c_{\text{meas},\hat{\imath}}(f_j)|^2 }{\langle|c_{\text{meas},\hat{\imath}}(f_j)|^2\rangle_{\hat{\imath}}} \right\rangle_{\hat{\imath},j} \quad.
\end{equation}
Note, that the normalization is based on the measured CSM and thus corresponds to the maximum observed source level per frequency, combined for all sources. Thus, this normalization does not change the relative amplitude levels between several sources at a given frequency, it only changes the combined source levels over frequency so that the energy is weighted equally at each frequency. We will refer to GO with eq.~\ref{eq:cost_function} for a single frequency (essentially equal to eq.~\ref{eq:E_standard_GO}) as standard GO, and eq.~\ref{eq:cost_function} for multiple frequencies as broadband GO. \deleted[R2C04]{To make the error more robust for real-world applications, and since eq.~\ref{eq:forward_operator} is no longer linear, it is also possible to choose the $L_1$ norm over the $L_2$ norm (which is typically chosen to ensure a convex error) which does not prefer single large CSM deviations in the optimization process. A weighted norm, based on the standard deviation of the CSM Welch averages~\citep{Raumer2021}, may further improve the estimation. To limit the several options, we will only analyze the formulation~\ref{eq:cost_function} as a baseline for broadband GO in this paper.}

\subsection{Energy of a single synthetic monopole}\label{sec:syn_monopole}

\begin{table}[]
    \centering
    \begin{tabular}{c|clllll}
        & $N_\text{true}$ & datatype & sourcetype& Section&Figures\\
        \hline
        case 1 & 1 &syn. & monopole&\ref{sec:syn_monopole}&\ref{fig:Figure01}, \ref{fig:Figure02}, \ref{fig:Figure03}, \ref{fig:Figure04}\\
        case 2 & 2 &syn.& monopole&\ref{sec:synth_error_2_sources}&\ref{fig:Figure05}, \ref{fig:Figure06}, \ref{fig:Figure07}\\
        case 3 & 1 &syn.& monopole&\ref{sec:GO_synthetic}&\ref{fig:Figure08}, \ref{fig:Figure09}, \ref{fig:Figure10}\\
        case 4 & 2 &syn.& monopole&\ref{sec:GO_synthetic}&\ref{fig:Figure08}, \ref{fig:Figure09}, \ref{fig:Figure10}\\
        case 5 & 3 &real& monopole&\ref{sec:res_GO_real_mono}&\ref{fig:Figure14},\ref{fig:Figure15},\ref{fig:Figure16}\\
        case 6 & 2 &syn.& multipole&\ref{sec:res_synth_dipole}&\ref{fig:Figure17}\\
    \end{tabular} 
    \caption{Main cases considered in this paper and their differences, e.g. the number of true sources $N_\text{true}$. The cases typically contain sub-cases, where depending on the employed method the properties $N_\text{est.}$, broad- and small-band energy, and Mach number are varied.}
    \label{tab:cases}
\end{table}

\begin{figure}[h!]
    \centering
    \includegraphics[width=3.48in]{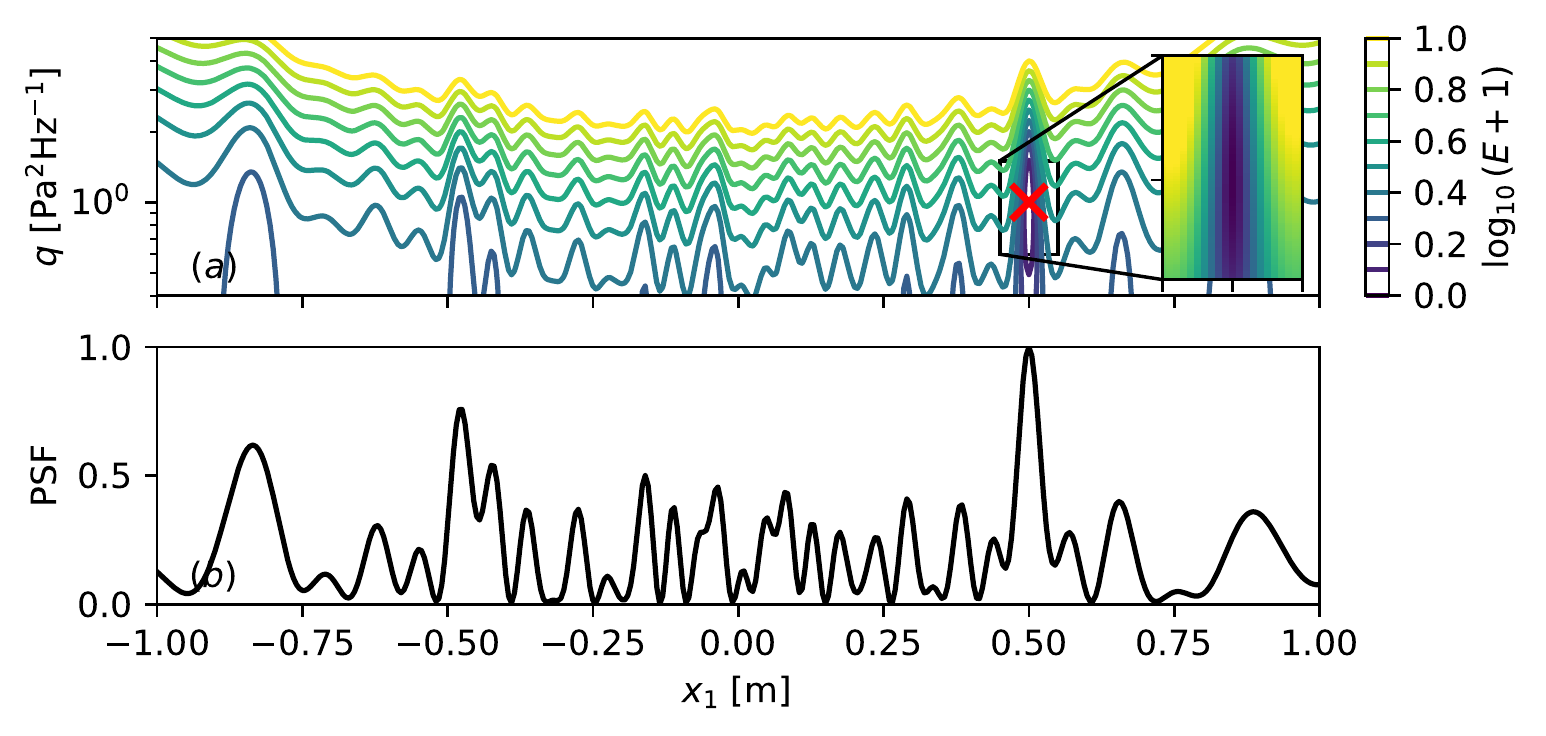}
    \caption{(Color online) Case~$1a)$. $(a)$ shows isocontour lines of the energy for ($x_1$,$q$) variations from eq.~\ref{eq:cost_function} for $x_2=\SI{0.5}{\metre}$, $x_3=\SI{0}{\metre}$, and $f=\SI{6144}{\hertz}$. The true source position is marked (red x). $(b)$ shows the corresponding PSF (steering-vector IV~\citep{Sarradj2012,Lehmann2022,Chardon2022b}) for a source location at $y=[0.5,0.5,0]^T$m.}
    \label{fig:Figure01}
\end{figure}
We will explore the energy numerically, due to the nonlinearity of the propagation operator, the non-convex energy function, and non-uniqueness (the solution is at least permutation invariant). This paper will contain several different setups, which are summarized in Table~\ref{tab:cases}. First, a synthetic monopole at $y=[0.5,0.5,0]^T$m with $q(f)=\SI{1}{\pascal\squared\per\hertz}$ is generated. An equidistant 1D-array at $\SI{-0.5}{\metre}\le x_1\le \SI{0.5}{\metre}$, $x_2=\SI{0}{\metre}$, $x_3=\SI{0}{\metre}$ detects the sound-field with $M=5$ microphones. We calculate the broadband GO energy from eq.~\ref{eq:cost_function} for a single frequency (equivalent to standard GO, see eq.~\ref{eq:E_standard_GO}) as case~$1a)$ at $f=\SI{6144}{\hertz}$, and for multiple frequencies as proposed in this paper at $2^{10}\si{\hertz}\le f\le2^{15}\si{\hertz}$ with $\Delta f=\SI{1024}{\hertz}$ as case~$1b)$. Note, due to the normalization over frequency in eq.~\ref{eq:cost_function}, the broad-band energy in case $1b)$ will be identical for frequency-dependent source amplitudes.

\begin{figure}[h!]
    \centering
    \includegraphics[width=3.48in]{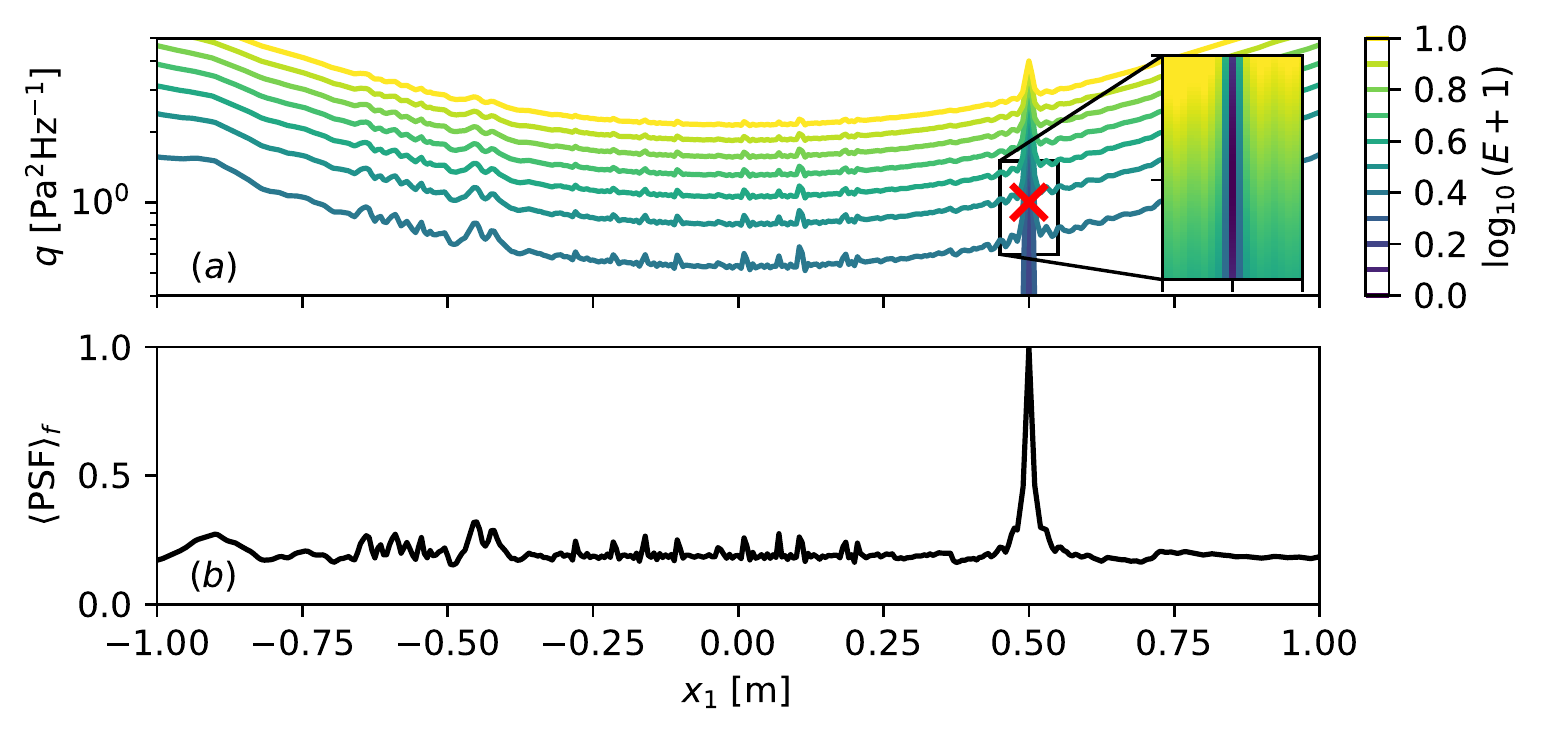}
    \caption{(Color online) Case~$1b)$. $(a)$ shows the energy for ($x_1$,$q$) variation from eq.~\ref{eq:cost_function} for $x_2=\SI{0.5}{\metre}$, $x_3=\SI{0}{\metre}$, and $\SI{100}{\hertz}\le f\le\SI{20}{\kilo\hertz}$, $\Delta f=\SI{100}{\hertz}$. $(b)$ shows the corresponding PSF (steering-vector IV~\citep{Sarradj2012,Lehmann2022,Chardon2022b}), averaged over the frequency, for a source location at $y=[0.5,0.5,0]^T$m.}
    \label{fig:Figure02}
\end{figure}
Figure~\ref{fig:Figure01} $(a)$ shows the case~$1a)$ log-energy for the variation of ($q$, $x_1$) for $x_2=\SI{0.5}{\metre}$, $x_3=\SI{0}{\metre}$. The variables that are not shown in the energy maps (i.e. $x_2$,$x_3$) are set correctly so that the energy deviations from zero are only caused by the displayed variables (i.e. $x_1$,$q$). The global minimum is at $E=0$, where the variables coincide with the true monopole location and source strength. Multiple local minima are visible for case~$1a)$, so that for a given $q$ the energy looks like a reciprocal Point Spread Function (PSF) with \replaced[R2C12]{grating}{side} lobes, shown in Figure~\ref{fig:Figure01} $(b)$. With the averaging over the frequency in case $1b)$, the PSF and local minima are also averaged, see Figure~\ref{fig:Figure02} $(a)$ and $(b)$. Thus, the energy becomes mostly smooth.

\begin{figure}[h!]
    \centering
    \includegraphics[width=3.48in]{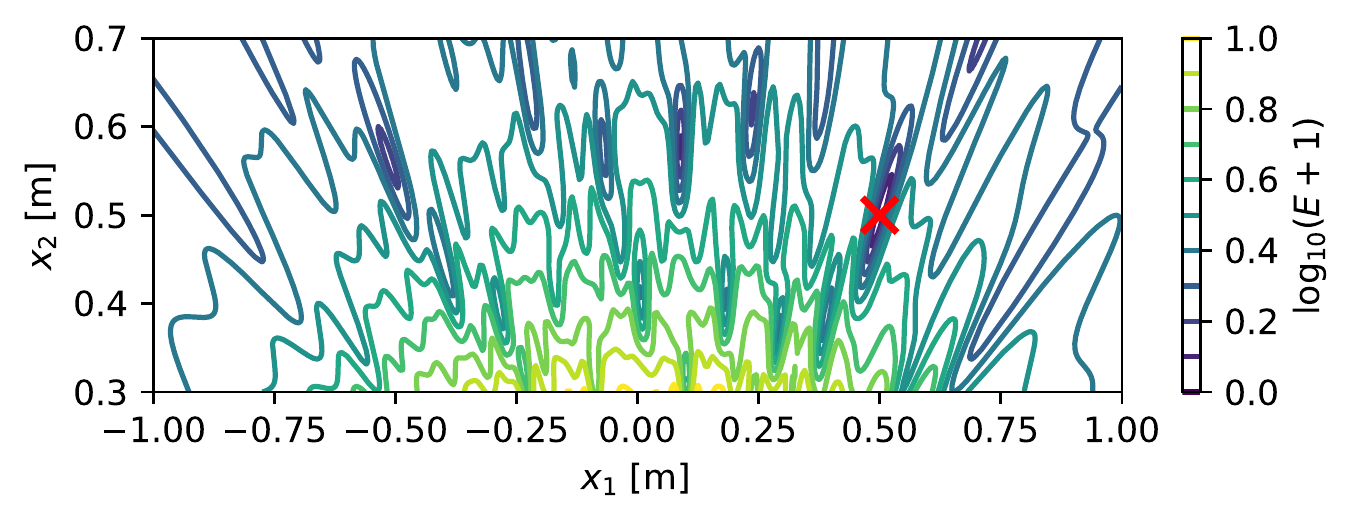}
    \caption{(Color online) Case~$1a)$, energy for ($x_1$,$x_2$) variation from eq.~\ref{eq:cost_function} for $x_3=\SI{0}{\metre}$, $q=\SI{1}{\pascal\squared}$, and $f=\SI{6144}{\hertz}$.}
    \label{fig:Figure03}
\end{figure}
\begin{figure}[h!]
    \centering
    \includegraphics[width=3.48in]{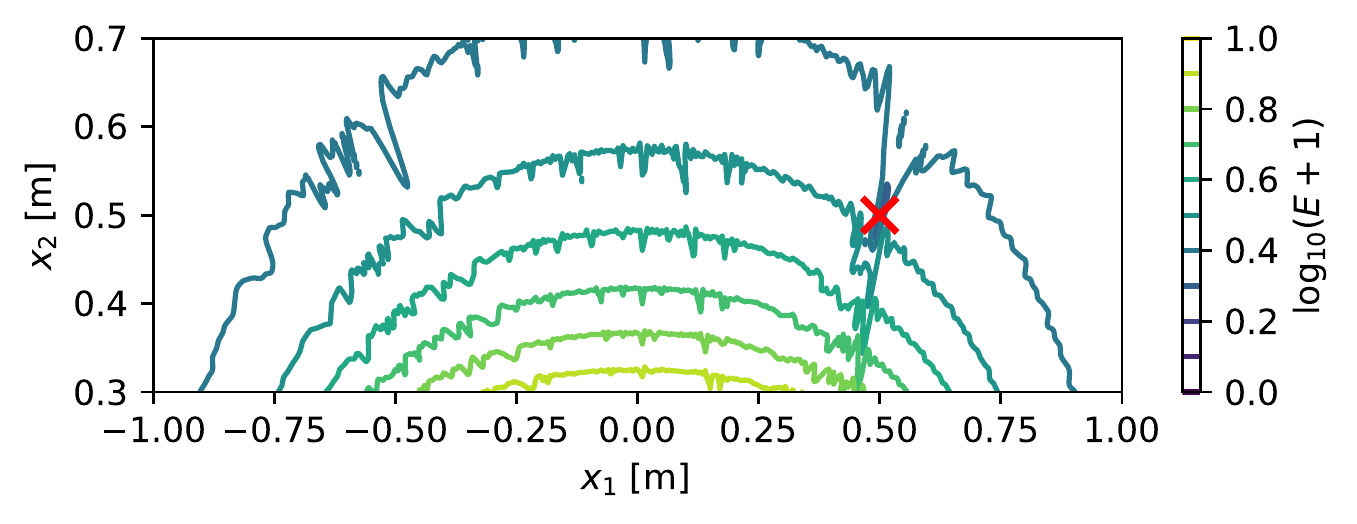}
    \caption{(Color online) Case~$1b)$, energy for ($x_1$,$x_2$) variation from eq.~\ref{eq:cost_function} for $x_3=\SI{0}{\metre}$, $q=\SI{1}{\pascal\squared}$, and $2^{10}\si{\hertz}\le f\le2^{15}\si{\hertz}$.}
    \label{fig:Figure04}
\end{figure}

Figure~\ref{fig:Figure03} and Figure~\ref{fig:Figure04} show the corresponding energy of variations of ($x_1$,$x_2$) for the correct amplitude. Again, most of the local minima of the single-frequency energy are smoothed for the broadband energy. Additionally, the gradient towards to global minimum is much steeper for the broadband energy. The depicted energy map will increase in smoothness with an increasing number of frequencies (either due to faster sampling rates or larger Welch block sizes, not shown here).

\subsection{Energy of multiple synthetic monopoles}\label{sec:synth_error_2_sources}
We will now examine the energy function for a two-source problem. The true sources will be denoted with Roman numbers. The synthetic monopoles are located at $y_{I}=[0.5,0.5,0]^T$m, $y_{II}=[0.5,0.6,0]^T$m, $q_{I}(f)=\SI{1}{\pascal\squared\per\hertz}$, $q_{II}(f)=\SI{0.5}{\pascal\squared\per\hertz}$, so that source $S_{II}$ is located slightly behind source  $S_{I}$. Again, the energy will be calculated for a single frequency $f=\SI{6144}{\hertz}$ as case $2a)$ and for multiple frequencies $2^{10}\si{\hertz}\le f\le2^{15}\si{\hertz}$ with $\Delta f=\SI{1024}{\hertz}$ as case $2b)$. Since the energy space is six-dimensional (three coordinates per source, one amplitude per source, because the amplitude is constant over frequency), we show only 2D ($x_1$,$x_2$)-slices of the energy-space. The slices are chosen so that all variables are correct except for the two being displayed. 

\begin{figure}[h!]
    \centering
    \includegraphics[width=3.48in]{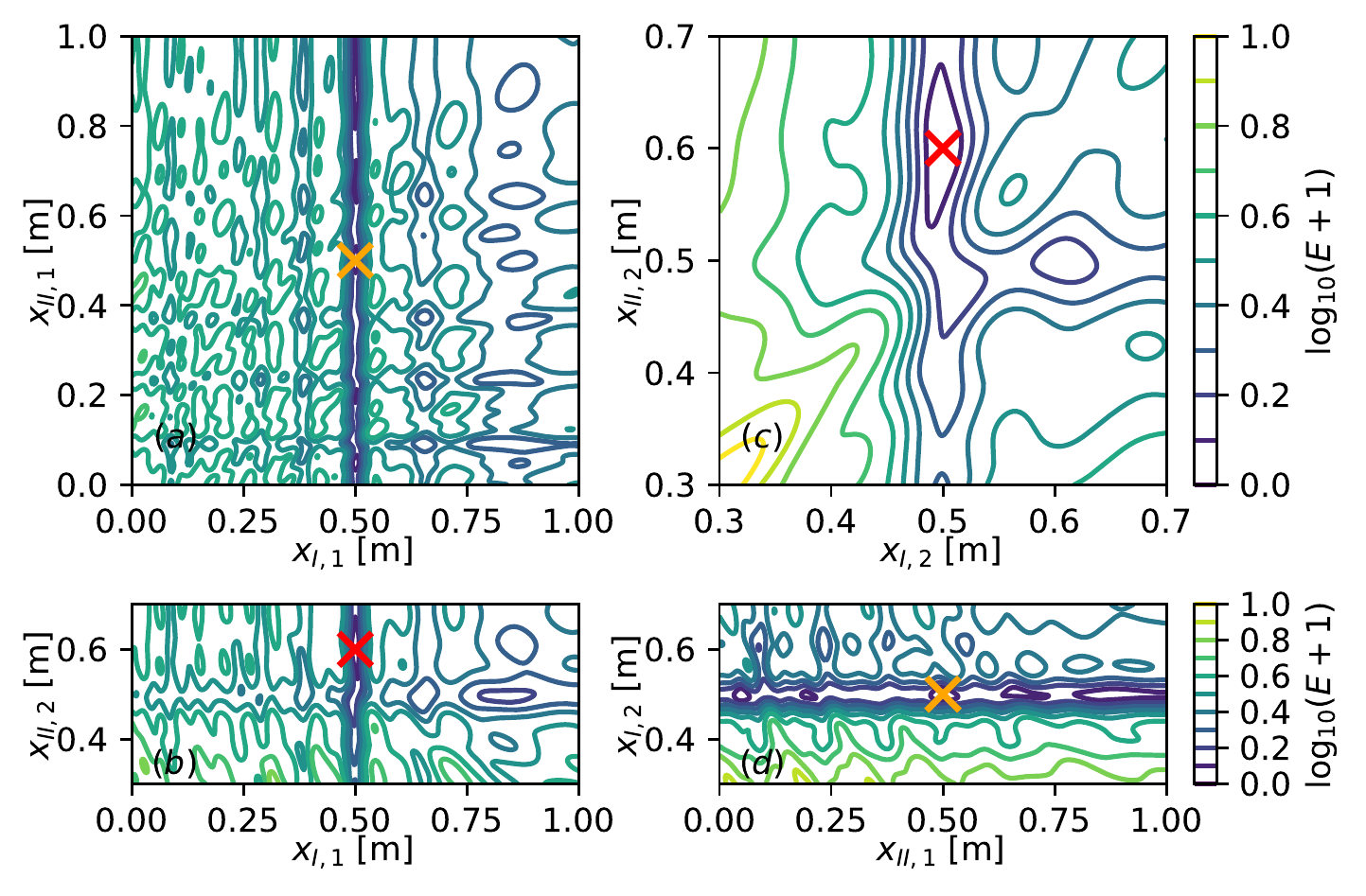}
    \caption{(Color online) Case $2a)$, slices of the six-dimensional energy, see eq.~\ref{eq:cost_function}. All variables for each slice are chosen correctly, except for the shown combinations of $(a)$ $(x_{I,1},x_{II,1})$, $(b)$ $(x_{I,1},x_{II,2})$, $(c)$ $(x_{I,2},x_{II,2})$, $(d)$ $(x_{II,1},x_{I,2})$. The source positions are marked (x) for $S_{I}$ (orange), and $S_{II}$ (red).}
    \label{fig:Figure05}
\end{figure}

\begin{figure}[h!]
    \centering
    \includegraphics[width=3.48in]{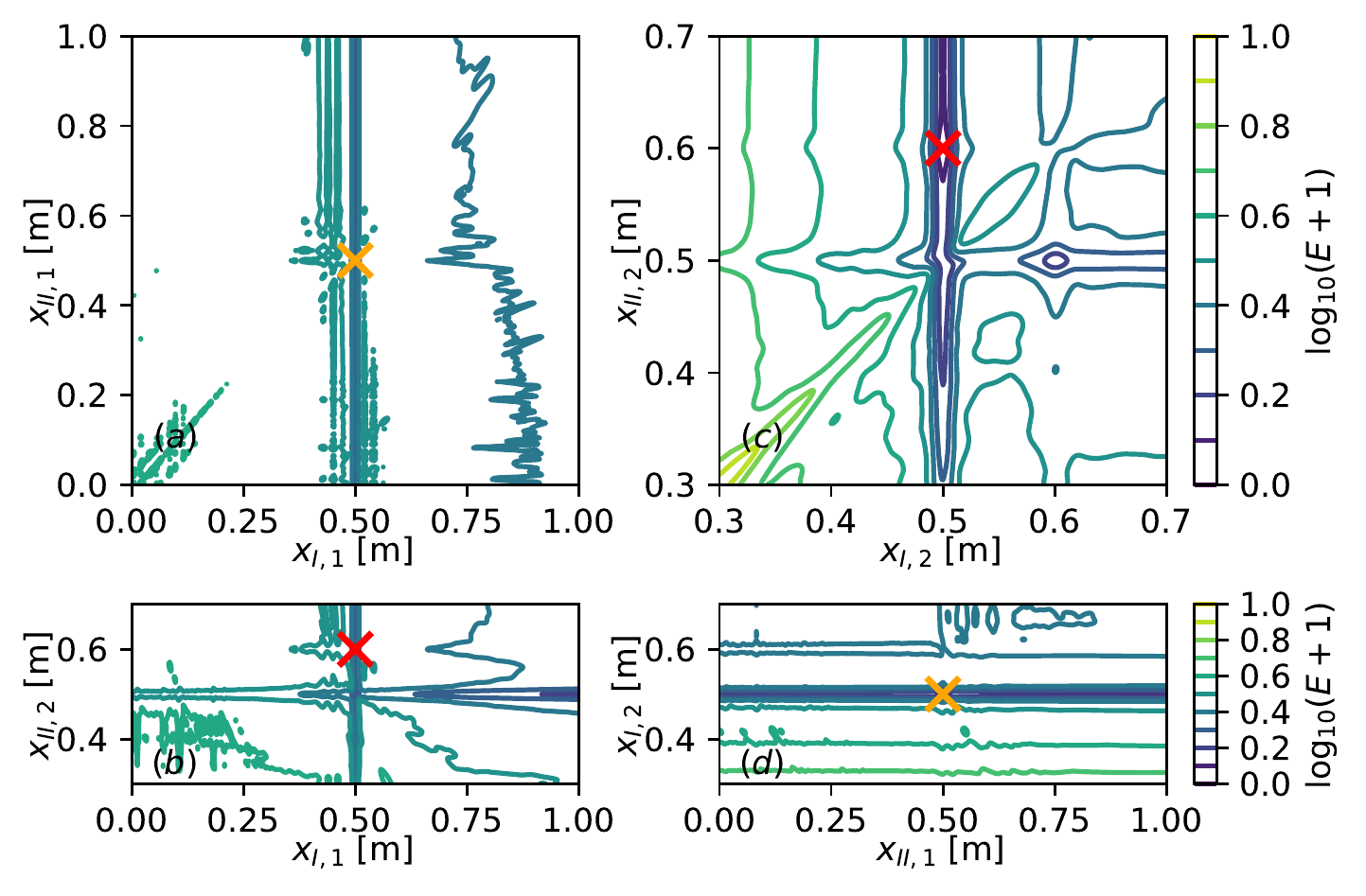}
    \caption{(Color online) Case $2b)$, slices of the six-dimensional energy, see eq.~\ref{eq:cost_function}. All variables for each slice are chosen correctly, except for the shown combinations of $(a)$ $(x_{I,1},x_{II,1})$, $(b)$ $(x_{I,1},x_{II,2})$, $(c)$ $(x_{I,2},x_{II,2})$, $(d)$ $(x_{II,1},x_{I,2})$. The source positions are marked (x) for $S_{I}$ (orange), and $S_{II}$ (red).}
    \label{fig:Figure06}
\end{figure}

Figure~\ref{fig:Figure05} shows the single frequency energy from eq.~\ref{eq:cost_function} for case~$2a)$. The true source positions are marked with an x. For the single frequency energy map $(a)$, $(b)$, and $(d)$, there is a trend of low energy at the correct $x_1$ positions visible. However, periodically repeating local minima are visible due to the array's PSF at the given frequency. Figure~\ref{fig:Figure06} shows the corresponding broadband energy for case $2b)$, which, in comparison, shows a clear global minimum with steep gradients. However, the energy map is not perfectly smooth and does exhibit minor local minima. Note, that these minima appear magnified due to the logarithmic plot of the energy. Figure~\ref{fig:Figure05} and Figure~\ref{fig:Figure06} $(c)$ show the variation of $x_2$ for both sources, for the correct $x_1=0.5$. Here, one can see that the main local minimum is caused by the spatial permutation of the sources (when $S_I$ and $S_{II}$ swap their positions), which again shows that the problem does not have a unique solution.
These results show, that the energy function, which is minimized during the GO CSM-fitting optimization, contains local minima, caused by the array's PSF. These minima can be reduced using multiple frequencies at once with the broadband energy formulation in eq.~\ref{eq:cost_function} so that the energy function is mostly smooth.

\begin{figure}[h!]
    \centering
    \includegraphics[width=3.48in]{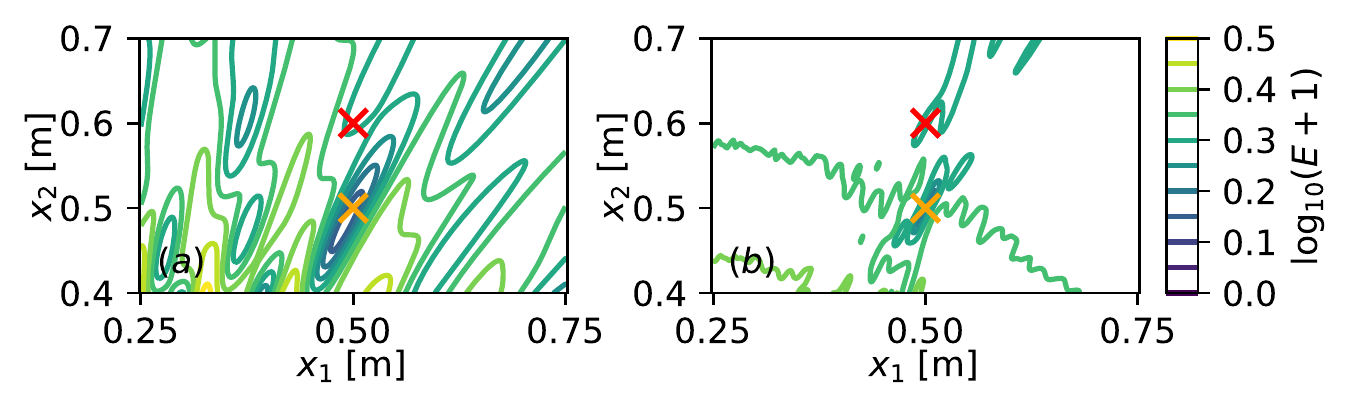}
    \caption{(Color online) $(a)$ smallband energy at $f=\SI{6144}{\hertz}$ and $(b)$ broadband energy of case $2c)$, with a mismatch of $N_\text{true}=2$, and $N_\text{est.}=1$ for $q=\SI{0.5}{\pascal\squared\per\hertz}, x_3=\SI{0}{\metre}$. }
    \label{fig:Figure07}
\end{figure}
A question that remains is if an estimated source is guaranteed to approximate a single true source, or if a single estimated source may be located between two real sources, trying to approximate both. This question is of interest since all estimated sources are fitted at once so that the errors of the estimated sources are not independent. While it seems that this is the case from Figure~\ref{fig:Figure05} and Figure~\ref{fig:Figure06}, where there is no strong local minimum between two true source positions, it is hard to tell, since we only visualize slices of the high-dimensional energy function. 

To explore this topic further, we will consider a case $2c)$, that has a mismatch of the true number of sources $N_\text{true}=2$, and estimated sources $N_\text{est.}=1$. Figure~\ref{fig:Figure07} $(a)$ shows the resulting asingle frequency energy, and Figure~\ref{fig:Figure07} $(b)$ the corresponding broadband energy. The energy is reduced for an estimation of either true source position, however, due to the closer distance of $S_I$ to the array and the decay in amplitude over distance, see eq.~\ref{eq:monopole_green}, an estimation of $S_I$ reduces the energy significantly more than an estimation of $S_{II}$. For both energies, the gradient around the source positions leads towards the true positions and again, there is no local minimum between the source positions. While this observation might be different for coherent and distributed sources, this means for \replaced[R2C10]{spatially compact}{point-like}, incoherent monopoles that each estimated source will approximate a single true source. If a local minimum is found for a source, the global minimum for other sources is at the correct location, see Figure~\ref{fig:Figure05} and Figure~\ref{fig:Figure06}

A mismatch between the number of true and estimated sources is no problem, and either an iterative addition of estimated sources~\citep{Chardon2023} until the energy is not reduced anymore, or a simple overestimation of sources will be a viable strategy to find all true sources.

\subsection{Source identification and spectra generation}\label{sec:ROI}
An additional problem that all beamforming methods that calculate results independently for each frequency share is that generating a spectrum is not trivial. The first problem is, that it is typically unknown how many sources are located within the map, how they are distributed, and in which frequency ranges they can be observed. The second problem is, that true sources are masked by beamforming artifacts such as side-lobes of other sources. The third problem is how to generate the correct spectrum from the beamforming map.

There are beamforming methods that produce dense maps, which means that in the observed region each focus point contains a source strength $q(y,f)>\SI{0}{\pascal\squared\per\hertz}$ such as conventional beamforming (also often referred to as a dirty map~\citep{MerinoMartinez2019a}). Here, the source identification and spectra generation problem is particularly difficult. Since the map is a convolution of the real sources with the array's PSF, the PSF's effect on the map's level has to be taken into account when integrating the map~\citep{Brooks1999}. While side-lobes can be partially suppressed by summing the maps over frequency~\citep{Simons2017} (e.g., third-octave bands), this reduces the frequency resolution. Even when summing all frequencies to obtain an Overall Sound Pressure Level (OASPL), the map is still three-dimensional and difficult to visualize and analyze.

Some methods produce sparse maps, which means that in a given region only a few locations contain a source strength such as CLEAN-SC~\cite{Sijtsma2007}, GO~\citep{Malgoezar2017,Hoff2022}, or machine learning-based methods~\citep{Kujawski2022}. We call these source energies $q(y,f)>\SI{0}{\pascal\squared\per\hertz}$ at single locations and single frequencies source-parts~\cite{Goudarzi2021}. Here, one can use this sparseness property to identify source positions based on the statistical spatial occurrence of source-parts directly using clustering such as K-Means~\citep{Kujawski2022}, HDBSCAN~\citep{McInnes2017,Goudarzi2021} (also including the frequency, and source strength information), or Gaussian Mixture Models~\cite{Goudarzi2021}. For simple problems with few sources, many results rely on manually defined Regions Of Interest (ROI). These are spatial regions that define a source and assign each source-part within them to the source. When two ROI spatially overlap, a source-part is typically assigned to the source with the closest distance to the ROI midpoint. Source-parts that are not located in any ROI are rejected as background noise.

All of these methods have in common that spectra are obtained by estimating the number and locations of the sources, assigning the source-parts either to a source or rejecting them as noise, and then integrating all source-parts per frequency that were assigned to a source.

For the proposed broadband GO method this procedure will be different since the source objects already include a full spectrum. However, multiple source objects may be located very close together, either because they approximate a distributed source with multiple point-like source objects, or because the solution of GO is not unique. Due to the super-position invariance of the CSM-fitting process, any number of source objects may be located at the same location, and the true source strength may be arbitrarily distributed between the estimated source objects. To account for these cases, we propose to define a minimum spatial distance, below which source objects will be grouped, and their spectra will be integrated. This is only valid for \replaced[R2C10]{spatially compact}{point-like} monopole sources, as more complex properties such as a dipole rotation will have to be taken into account when integrating sources.

\section{Results}\label{sec:results}
This section presents standard GO, broadband GO, and CLEAN-SC results for increasingly difficult problems, such as synthetic monopoles, real monopoles, and synthetic dipoles, listed in Table~\ref{tab:cases}. For GO, many algorithms exist, that feature a variety of hyper-parameters that can be tuned~\cite{Malgoezar2017,Hoff2022}. Since the main goal is to establish a baseline of broadband GO, hyper-parameter tuning and an exploration of different GO algorithms are out of the scope of this paper. Throughout the paper, we will use dual annealing~\citep{2020SciPy-NMeth} with SciPy's standard hyper-parameters. For local optimization, we will use L-BFGS-B~\citep{2020SciPy-NMeth} with SciPy's standard hyper-parameters.

\subsection{Global optimization on synthetic monopoles}\label{sec:GO_synthetic}
To evaluate the proposed method, and to compare it to standard GO and CLEAN-SC~\citep{Sijtsma2007} we propose two examples. First, case~$3)$ contains a monopole located at $y_{I}=[0.5,0.5,0]^T$m, with $Q_{I}(f)=\SI{100}{\decibel\per\hertz}$, where $Q=10\log_{10}\left(q/\SI{4e-10}{\pascal\squared}\right)$. Second, case~$4)$ contains two incoherent monopoles with $y_I=[0.5,0.5,0]^T$m, $Q_I(f)=a_0f+b_0\SI{}{\decibel\per\hertz}$ with $a_0 = 5/7936$ and $b_0 = 2770/31$ so that it increases linearly from $Q_I(2^{10} \SI{}{\hertz})=\SI{90}{\decibel\per\hertz}$ to $Q_I(2^{15} \SI{}{\hertz})=\SI{110}{\decibel\per\hertz}$, and $y_{II}=[0.5,0.6,0]^T$m, $Q_{II}=\SI{100}{\decibel\per\hertz}$. Thus, source $S_{II}$ is located behind source $S_{I}$ (similar to Sec.~\ref{sec:synth_error_2_sources}). The array is an equidistant 1D-array at $\SI{-0.5}{\metre}\le x_1\le \SI{0.5}{\metre}$, $x_2=\SI{0}{\metre}$, $x_3=\SI{0}{\metre}$ with $M=11$ microphones. To acquire depth-information with conventional beamforming, we choose steering vector formulation IV~\citep{Sarradj2012,Lehmann2022,Chardon2022b} and a focus grid $\SI{-1}{\metre}\le x_1\le\SI{1}{\metre}$, $\SI{0.3}{\metre}\le x_2\le\SI{0.7}{\metre}$ with $\Delta x_1=\Delta x_2=\SI{0.01}{\metre}$, and $x_3=\SI{0}{\metre}$.

\begin{figure}[h!]
    \centering
    \includegraphics[width=3.48in]{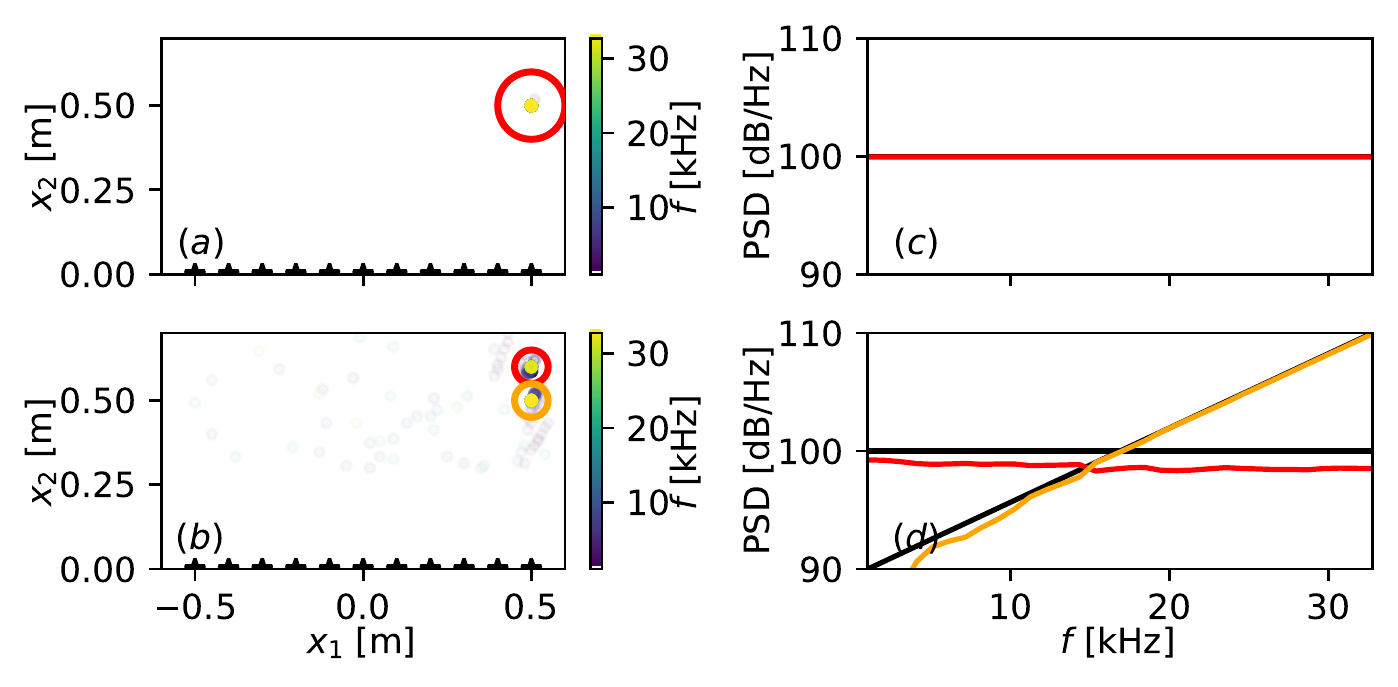}
    \caption{(Color online) CLEAN-SC result for top row case~$3)$, bottom row case~$4)$. $(a)$ and $(b)$ show the spatial distribution of microphones ($\star$), and of the source-parts ($\cdot$), their color encodes the corresponding frequency. The opacity indicates the per frequency normalized log-power. $(c)$ and $(d)$ show the integrated spectra, based on the corresponding ROI (o) in $(a)$ and $(b)$. The ground truth PSD is depicted in black.}
    \label{fig:Figure08}
\end{figure}

Figure~\ref{fig:Figure08} shows the corresponding CLEAN-SC result, top row case~$3)$, bottom row case~$4)$. The first column shows the spatial position of the reconstructed source-parts. A source-part is defined as a single $q$ in a single location and frequency with $q(x,f)>\SI{0}{\pascal\squared\per\hertz}$, that - once integrated through a ROI - provides a source spectrum~\citep{Goudarzi2021}, see Sec.~\ref{sec:ROI}. The source-parts' color indicates their frequency, and their opacity indicates their log power-level $\log(q+1)$, normalized for each frequency independently to $(0,1)$. In Figure~\ref{fig:Figure08} $(a)$ all source-parts are (correctly) located at the same position, thus, only one source-part is visible. The ROI are manually defined at the true source locations with case~$3)$ radii $r_I=\SI{0.1}{\metre}$, and case~$4)$ $r_{II}=\SI{0.05}{\metre}$ so that the ROI do not overlap. Figure~\ref{fig:Figure08} shows in the right column the corresponding integrated spectra. Source-parts that lie outside of any ROI are integrated and labeled as (background) noise. The position and PSD of the single monopole configuration in case~$3)$ are estimated perfectly. For case $4)$ the dominant source-parts are located close to the true source position. However, there is a lot of noise at low frequencies, and the PSD of source $S_{II}$ is underestimated.

For GO we approach the solution with sub-cases. GO requires a fixed number of estimated sources, which is typically not known beforehand. To be able to find and resolve all sources, a simple strategy is to overestimate the number of sources, see Sec.~\ref{sec:synth_error_2_sources}. Thus, we first perform standard GO with the correct number of estimated sources $N=1$ as case~$3a)$, then we perform standard GO with an additional estimated source $N=2$ as case~$3b)$. For case~$4)$ we will underestimate $N_\text{est.}=1$ as case $4a)$ and overestimate $N_\text{est.}=3$ as case $4b)$. True sources will be denoted with Roman numbers (e.g. $S_I$),  and estimated sources with Arabic numbers (e.g. $S_1$).

\begin{figure}[h!]
    \centering
    \includegraphics[width=3.48in]{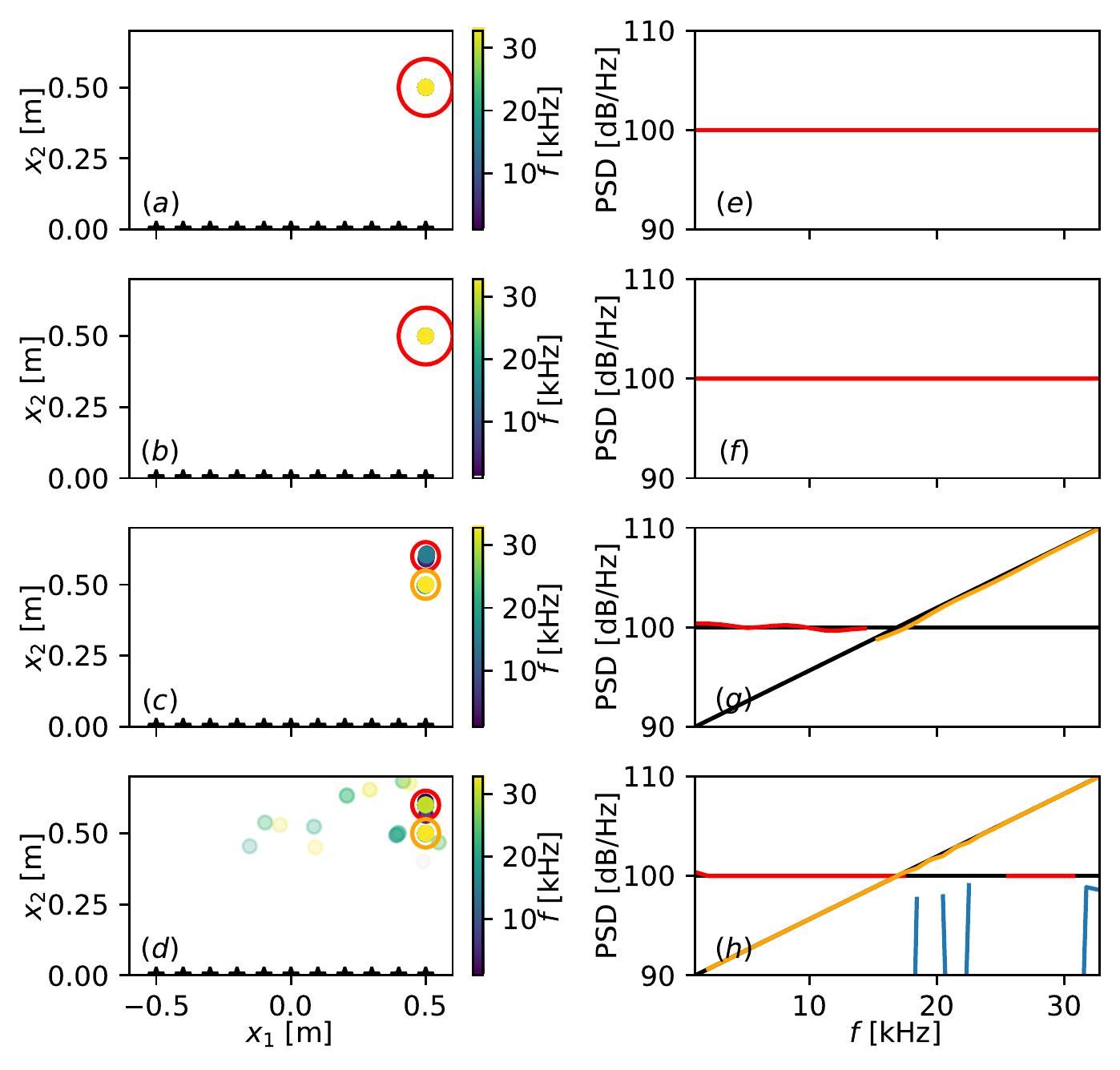}
    \caption{(Color online) Standard GO. First row: case~$3a)$, second row: case~$3b)$, third row: case~$4a)$, forth row: case~$4b)$. $(a)$, $(b)$, $(c)$, and $(d)$ show the estimated source locations ($\cdot$), the color displays the corresponding frequency, the opacity the per frequency normalized log-power. $(e)$, $f)$, $g)$, and $(h)$ show the PSD estimation (-), integrated from the corresponding ROI (o) in the left row. The ground-truth is depicted in black, sources that do not lie in any ROI are integrated as noise (blue).}
    \label{fig:Figure09}
\end{figure}

Figure~\ref{fig:Figure09} shows the results of standard GO. The left column shows the estimated position of the source-parts, the color encodes their frequency. For cases $3a)$, and $3b)$ the source-parts are located mostly in the same (correct) location, for case $4)$ the source-parts are scattered around the map at medium to high frequencies. Since for standard GO a source location is estimated for each frequency separately we also define ROI according to the CLEAN-SC result, see Sec.~\ref{sec:ROI}. The spectra are correctly reconstructed for cases $3a)$, and $3b)$. For case $4a)$ the single estimated source always (correctly) estimates the dominant source. For case $4b)$ the reconstruction fails for  $S_{II}(f\ge\SI{17}{\kilo\hertz})$, which is likely due to the spatial aliasing and local minima in the energy caused by the side- and grating lobes, see Sec.~\ref{sec:synth_error_2_sources}. Thus, one can expect the results to gradually improve by increasing the number of global optimization searches~\citep{Hoff2022}.

\begin{figure}[h!]
    \centering
    \includegraphics[width=3.3in]{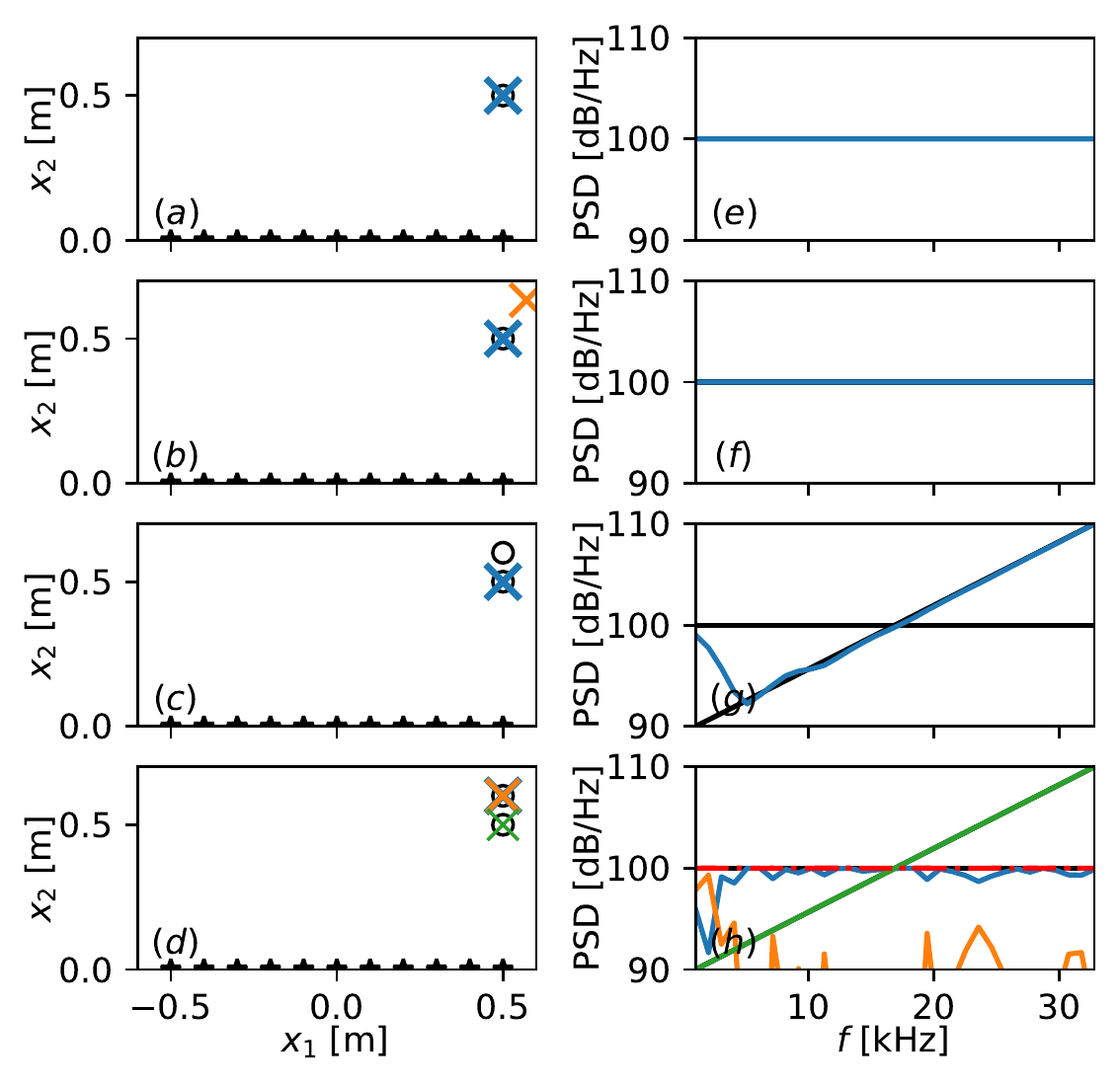}
    \caption{(Color online) Broadband GO. First row: case~$3a)$, second row: case~$3b)$, third row: case~$4a)$, forth row: case~$4b)$. $(a)$, $(b)$, $(c)$, and $(d)$ show the true source locations (o) and estimated source object locations (x), $(e)$, $(f)$, $(g)$, and $(h)$ show the corresponding PSD estimations (-). The dashed, red line represents the PSD summation of $S_1$ (blue) and $S_2$ (orange), which are located at the same spatial position.}
    \label{fig:Figure10}
\end{figure}

Figure~\ref{fig:Figure10} shows the result of the proposed broadband GO method. For case~$3a)$ and case $3b)$ the source location and amplitude is estimated correctly, like for standard GO, and CLEAN-SC. For case~$3b)$ $S_2$ is arbitrarily positioned with a low PSD and thus, can be neglected. For case $4a)$ $S_1$ spatially estimates the source with the highest overall contribution to the CSM ($S_I$), but at very low frequencies where the array cannot resolve $S_I$ and $S_{II}$ anymore, $S_1$ estimates $S_{II}$, which contributes more towards the CSM. For case~$4b)$ $S_{II}$ is estimated with the superposition of $S_1$ and $S_3$, since their position is identical, see Sec.~\ref{sec:ROI}. The spectral density is arbitrarily distributed between them, but the summation of both spectra reveals that their sum approximates the true PSD. The spectra are reconstructed well throughout the frequency range and the method outperforms CLEAN-SC and standard GO (for the given number of iterations). The results indicate, that an overestimation of sources is a viable strategy to find and reconstruct all true sources with low errors.

\subsection{Global optimization on real monopoles}\label{sec:res_GO_real_mono}
\begin{figure}[h!]
    \centering
    \includegraphics[width=3.4in]{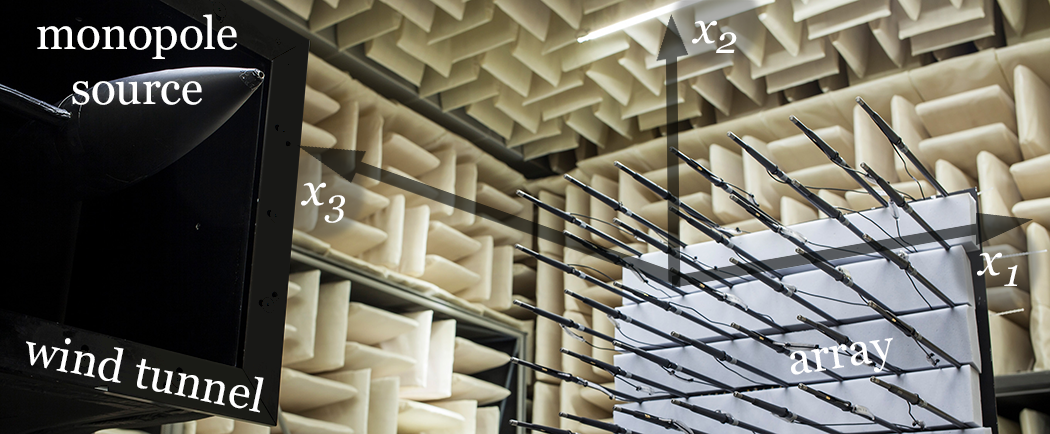}
    \caption{(Color online) Experimental setup, the monopole source is located in the core flow of the open wind tunnel section (left), the housing opening is at the downstream end (positive $x_1$ direction). On the right the equidistant 7x7 B\&K 4961 multi-field microphone array is mounted with 30cm long rods on a back-plate, that is then covered with 7cm Basotect absorbers. The coordinate system only indicates the directions, the actual coordinate origin lies roughly at the speaker housing opening.}
    \label{fig:Figure11}
\end{figure}

To evaluate the proposed method on real data, we use previously presented open wind tunnel data~\citep{Goudarzi2021} at Mach $\text{M}=0$ as case~$5a)$. The data features an equidistant 7x7 microphone array $-\SI{0.27}{\metre}\le x_{1,2}\le\SI{0.27}{\metre}$, $x_3=-\SI{0.65}{\metre}$, with $\Delta x_1=\Delta x_2 = \SI{0.09}{\metre}$, and a generic monopole source (streamlined housing with a circular $r=\SI{2.5}{\milli\metre}$ opening at the downstream end) in a mostly reverberation-free environment. The source is moved to the three locations 
\begin{equation}
    \begin{bmatrix} y_I\\y_{II}\\y_{III}\end{bmatrix}
    =
    \begin{bmatrix*}[r]
    -0.05&0.1&0.0\\
    0.10&0.1&0.0\\
    0.25&0.1&0.0
    \end{bmatrix*}
\end{equation}
during separate measurements and uses uncorrelated white noise with three different band-pass frequencies to generate different source spectra. Figure~\ref{fig:Figure11} shows the experimental setup.

The CSM has a sampling frequency of $f_S=2^{16}\si{\hertz}$, and uses a blocksize of $2^{6}$, which results in around $\SI{6e4}{}$ Welch averages and $\Delta f=2^{10}\si{\hertz}$. We estimate the ground truth source powers by dividing the CSMs by the propagation matrices $\textbf{T}'=\textbf{H}^\dag\textbf{H}$, see eq.~\ref{eq:Greens_matrix}, for an ideal monopole source, see eq.~\ref{eq:monopole_green}, where $\dag$ denotes the Hermitian (conjugate transpose). Then, we average the absolute, upper triangular CSM entries $\hat{\imath}$ to obtain an average ground truth spectrum with standard deviation.
\begin{equation}\label{eq:PSD_GT}
    \text{PSD}_{\text{true}}(f) = \left\langle \left|\frac{\textbf{C}(f)}{\textbf{T}'(f)}\right|\right\rangle_{\hat{\imath}}
\end{equation}
We then use a superposition of the three measurements (addition of their CSM) to obtain a single CSM that contains three independent sources. Note, that this source reconstruction problem is fairly difficult for beamforming, since the array's spatial resolution is low, and the array's PSF contains strong grating- and side lobes.

\begin{figure}[h!]
    \centering
    \includegraphics[width=3.48in]{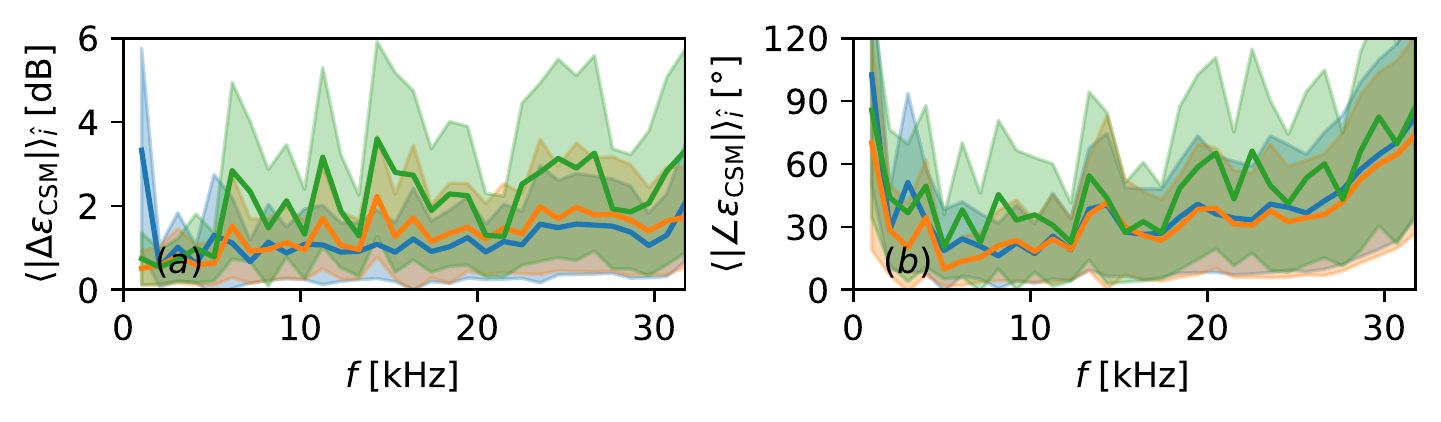}
    \caption{(Color online) $(a)$ shows the Mean Absolute Error (MAE) of the upper triagonal CSM entries ($\hat{\imath}$) between the synthetic and measured CSMs for the three source positions ($S_I$ depicted in blue, $S_{II}$ in orange, $S_{III}$ in green) with $1\sigma$ standard deviation (shaded area). $(b)$ shows the MAE of the corresponding phase difference.}
    \label{fig:Figure12}
\end{figure}

To evaluate how close the real sound source to an ideal monopole is, we calculate a synthetic CSM based on the average ground truth spectra, see eq.~\ref{eq:PSD_GT}, the true source locations, and the compact monopole assumption. We then calculate the Mean Absolute Error (MAE) of the upper triangular synthetic and measured CSM. Figure~\ref{fig:Figure12} shows the resulting error, which is calculated for the absolute PSD difference $(a)$, and absolute phase difference $(b)$. The errors show that the monopole assumption is reasonable at $\SI{1}{\kilo\hertz}\le f \le \SI{25}{\kilo\hertz}$. At lower frequencies the assumption is probably violated by reflections of the array's back-plate, at higher frequencies the assumption is gradually violated by the speaker design (non-compactness, asymmetric orientation towards the array). The influence of the asymmetric orientation towards the array can be also observed from the increased error from source position $y_I$ to $y_{II}$, and finally $y_{III}$. It will be of particular interest how the different methods will perform at these very low and high frequencies, where the method's assumptions are violated.

\begin{figure}[h!]
    \centering
    \includegraphics[width=3.48in]{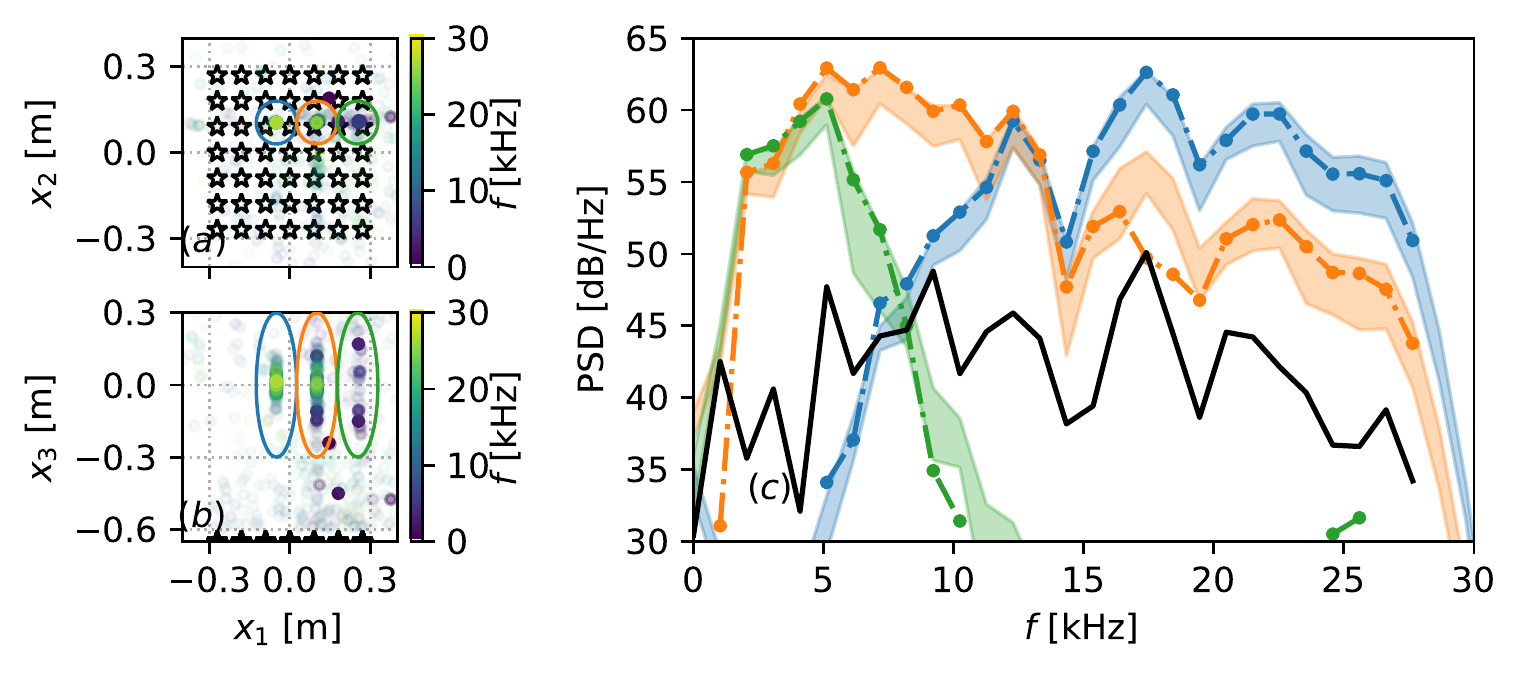}
    \caption{(Color online) CLEAN-SC results for case~$5a)$. $(a)$ and $(b)$ show projections of the spatial distribution of source-parts, their frequency, and normalized amplitude. $(c)$ shows the PSD reconstruction (dotted lines), based on same colored ROI with $r_1=r_2=\SI{0.075}{\metre}$, $r_3=\SI{0.3}{\metre}$. The ground truth is depicted in the same color with $\pm 1\sigma$ standard deviation. The black line indicates noise, that is the integration of the region that does not correspond to any ROI.}
    \label{fig:Figure13}
\end{figure}

We perform beamforming and CLEAN-SC with steering vector formulation IV~\citep{Sarradj2012,Lehmann2022,Chardon2022b} as a benchmark, see Figure~\ref{fig:Figure13}. The left column depicts the spatial setup and source-parts, their color indicates the frequency, their opacity their per-frequency normalized SPL. Since the problem is three-dimensional, $(a)$ shows a ($x_1,x_2$) and $(b)$ shows a ($x_1,x_3$) projection. There are three ROI defined based on the true source positions with a radius of $r_1=r_2=\SI{0.075}{\metre}$, $r_3=\SI{0.3}{\metre}$. Due to the low array resolution in $x_3$, the ROI are elongated with $r_3=4r_{1,2}$. Every source-part within these radii is integrated for the estimated source spectrum, and every source-part that is not contained in any ROI is integrated and classified as noise. Figure~\ref{fig:Figure13} $(c)$ shows the corresponding spectra, the estimated ROI spectra are depicted with dashed lines, and the ground truth is depicted as the mean $\pm 1\sigma$ standard deviation. Overall, CLEAN-SC is able to reconstruct the spectra up to a $\text{SNR}\approx\SI{30}{\decibel}$ at around $f\approx\SI{10}{\kilo\hertz}$. There is an underestimation of the source power of $S_2$ at $f\approx\SI{17}{\kilo\hertz}$. At high frequencies $f\ge\SI{28}{\kilo\hertz}$ the source reconstruction fails. Additionally, there is a lot of noise at all frequencies that does not correspond to any ROI. It is mainly located at the grating- and side lobe locations~\citep{Goudarzi2021}. Thus, an optimized array geometry would improve the results.

\begin{figure}[h!]
    \centering
    \includegraphics[width=3.48in]{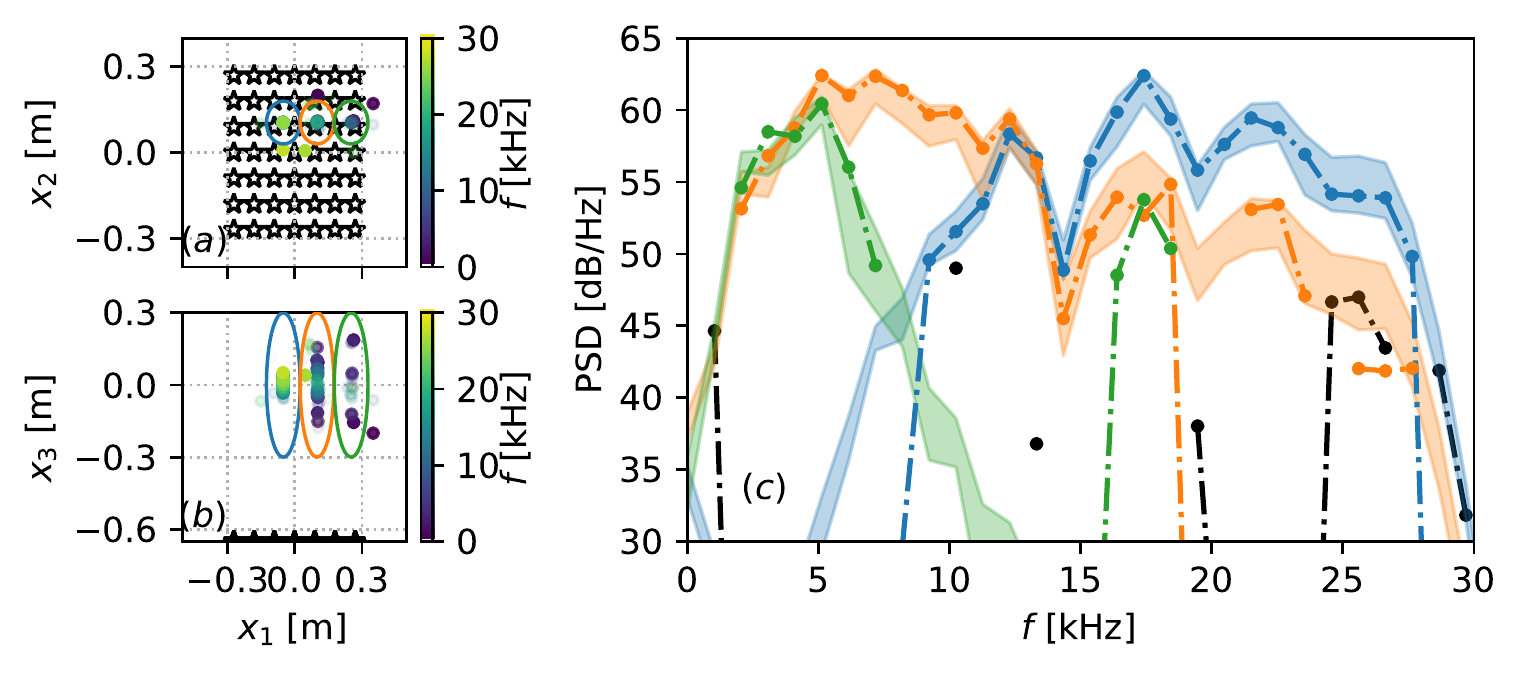}
    \caption{(Color online) Standard GO results for case~$5a)$, three true sources, and $N=4$. $(a)$ and $(b)$ show projections of the spatial distribution of source-parts, their frequency, and normalized amplitude. $(c)$ shows the PSD reconstruction (dotted lines), based on same colored ROI with $r_1=r_2=\SI{0.075}{\metre}$, $r_3=\SI{0.3}{\metre}$. The ground truth is depicted in the same color with $\pm 1\sigma$ standard deviation. The black line indicates noise, that is the integration of the region that does not correspond to any ROI.}
    \label{fig:Figure14}
\end{figure}

Figure~\ref{fig:Figure14} shows the standard GO results for $N=4$. The space-frequency source-part distribution is very similar to CLEAN-SC. While the reconstruction of $S_I$ at $f\approx\SI{17}{\kilo\hertz}$ is more accurate, CLEAN-SC outperforms standard GO at low frequencies and low SNR. Generally, the PSD estimation is good for dominant sources, but the locations of the source-parts are not well estimated so that they are often located in the wrong ROI. Again, we assume the results to improve with an optimized array geometry and an increasing number of global searches.

\begin{figure}[h!]
    \centering
    \includegraphics[width=3.48in]{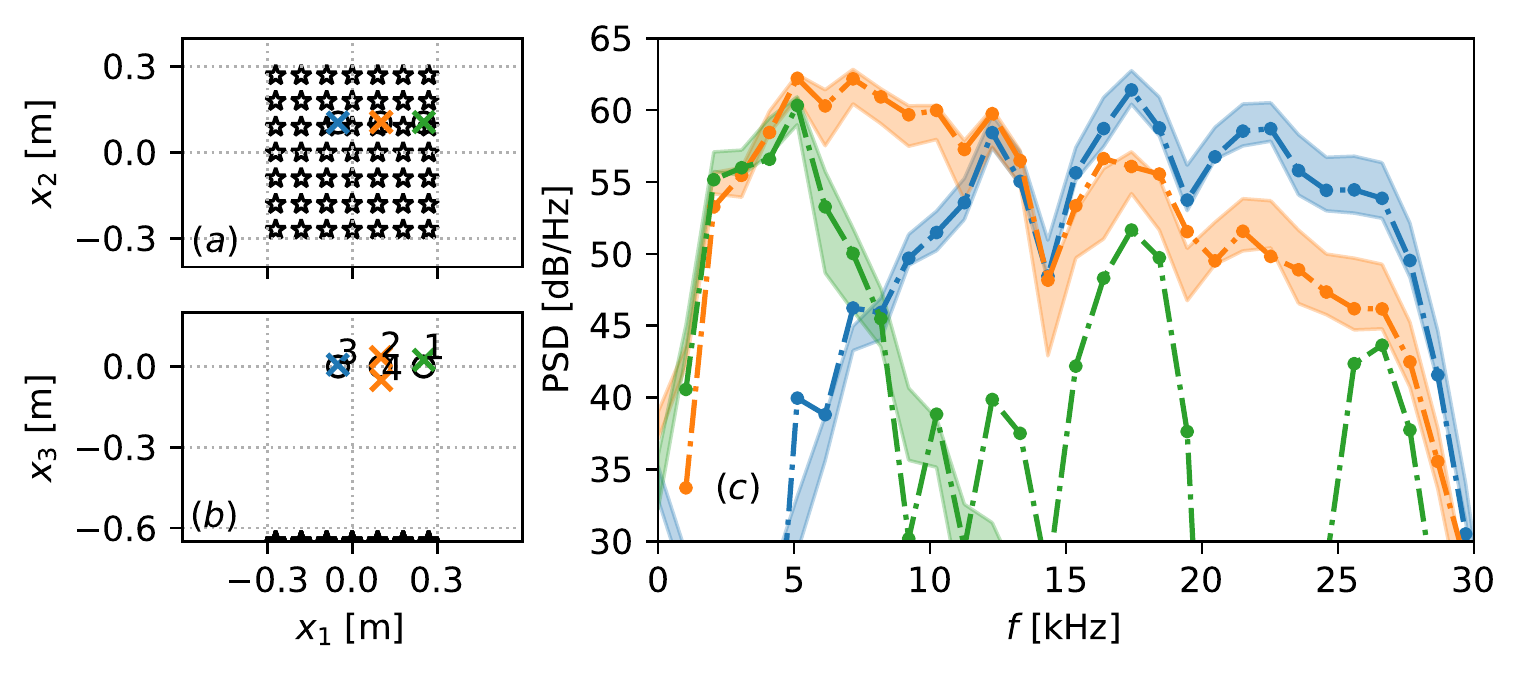}
    \caption{(Color online) Broadband GO result for case~$5a)$, three true sources, and $N_\text{est.}=4$. $(a)$ and $(b)$ show projections of the spatial distribution of the estimated source objects $S_i$, their ordering is shown in $(b)$. $(c)$ shows the PSD reconstruction (dotted lines), based on the single or integrated source objects ($S_{II}$ is approximated by $S_2$ and $S_4$). The ground truth is depicted in the same color with $\pm 1\sigma$ standard deviation.}
    \label{fig:Figure15}
\end{figure}

Figure~\ref{fig:Figure15} shows the result of the proposed broadband GO for $N=4$. On the left column, the estimated source positions are depicted, on the right the corresponding PSD is depicted. The estimated source object positions (x) are depicted in the same color as the true sources, $S_{II}$ (orange) is estimated by two source objects. Since the ordering of the sources is random, these correspondences were identified based on the spatial distances to the true positions. Note that this assignment problem only arises because of the comparison to the ground truth and does not need to be solved for typical beamforming applications. Similar to the CLEAN-SC result, the sources are well localized in $x_1$ and $x_2$, and $S_{II}$ is slightly distributed in $x_3$ direction. Figure~\ref{fig:Figure15} $(c)$ depicts the estimated source spectra by integrating all estimated source objects that correspond to a true source position, which correspond well to the ground source, except for $S_{III}$ which shows two prominent high-frequency bumps.

Overall, the CLEAN-SC and broadband GO results are very similar in terms of SNR, but broadband GO outperforms CLEAN-SC at $f\approx\SI{17}{\kilo\hertz}$ and high frequencies. The main difference is, that the CLEAN-SC result is very noisy, while there are two prominent bumps in the estimated spectrum of $S_{III}$. The similar shape of the CLEAN-SC noise and GO bumps are likely caused by the before-mentioned monopole assumption violation and reflections, as can be also seen for the same frequencies in Figure~\ref{fig:Figure12}.
This inclusion seems to minimize the observed CSM, similar to case $4a)$ in Figure~\ref{fig:Figure10} $(g)$, where at low frequencies the dominant source PSD is estimated instead of the true PSD.

\subsection{Local optimization on real monopoles}
Given we use a method like SIND~\citep{Goudarzi2021} to automatically obtain source positions and spectra from CLEAN-SC maps, we can use this information as initial guesses to transition from Global Optimization to Local Optimization (LO), since the error function is smooth in the amplitude dimension, see Section~\ref{sec:error}. The main reason to perform broadband GO or LO after acquiring a CLEAN-SC result would be to include properties in the source object that are neglected by conventional beamforming and CLEAN-SC, such as multipoles. The main reason to perform LO over GO is that GO is computationally very expensive~\citep{Neumaier2004,Chardon2023}.

\begin{figure}[h!]
    \centering
    \includegraphics[width=3.48in]{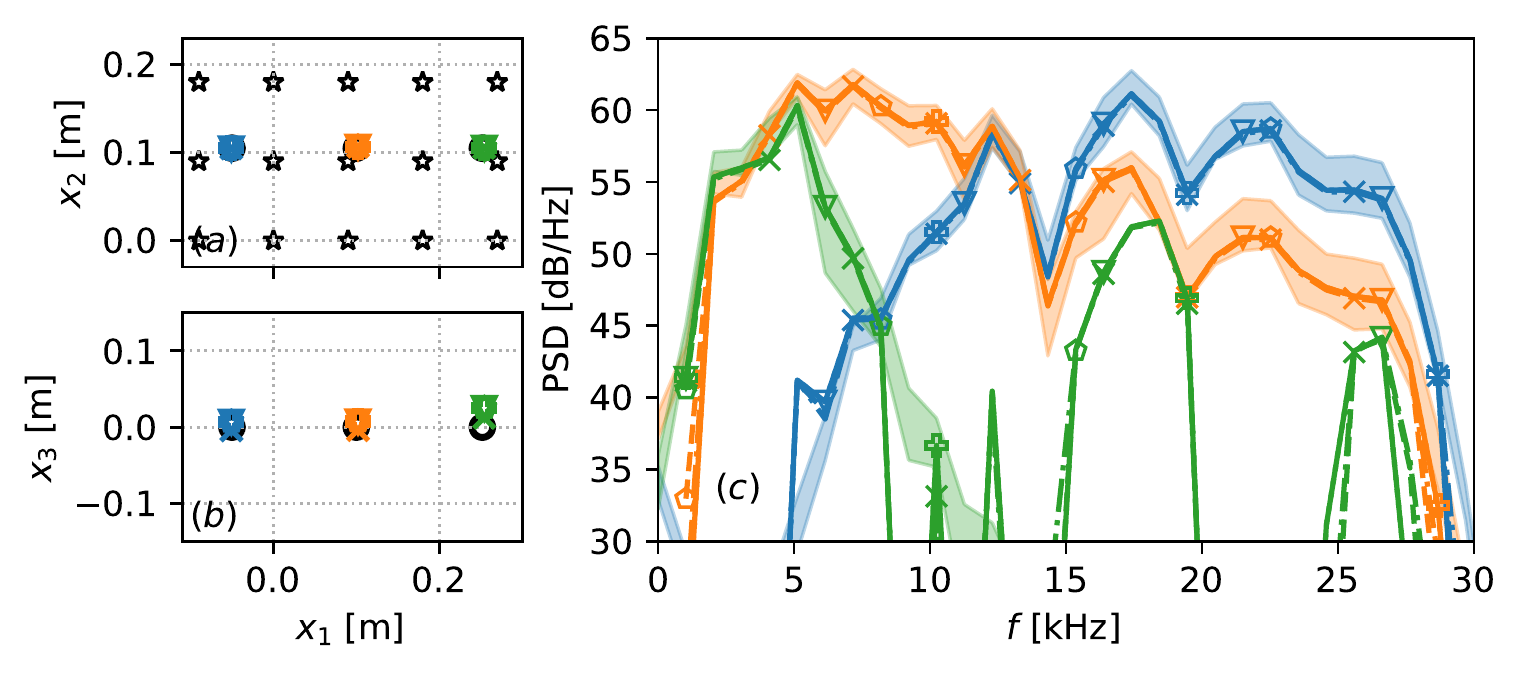}
    \caption{(Color online) Broadband LO result for case $5b)$. $(a)$ and $(b)$ show spatial projections of the estimated source objects, $(c)$ shows the PSD reconstruction for three true sources at $\text{M}=0.00$ (x, dash dotted line), $\text{M}=0.03$ ($\bigtriangledown$, dotted line), $\text{M}=0.06$ (+, dashed line), $\text{M}=0.12$ ($\pentagon$, solid line),  $N_\text{est.}=3$, with initial start positions close to the true positions, and $\text{PSD}_\text{start}=\SI{50}{\decibel}$.}
    \label{fig:Figure16}
\end{figure}

As an example, we perform LO on the same setup from case $5a)$, but with $\text{M}=[0.00,0.03,0.06,0.12]$ as case~$5b)$. The Amiet open wind tunnel correction~\citep{Amiet1976} is incorporated into the propagation operator for $\text{M}>0$. We use the true source positions with a random normal error of $\sigma_x=0.025$ as initial start values and bound the optimizer with $\pm 4\sigma_x$. The initial amplitude is $\text{PSD}_\text{start}=\SI{50}{\decibel}$, and not bounded. Figure~\ref{fig:Figure16} shows the result of LO. The results are nearly identical to broadband GO, and the Mach number has a neglectable influence on the outcome. This shows, that the proposed broadband energy is resistant to uncorrelated background noise (i.e. noise from the open wind tunnels). High background-noise situations such as closed wind tunnel measurements are likely to require a noise model~\citep{HaxterSpehr2012,Haxter2014,HaxterSpehr2016} either subtracted from the measurement or incorporated in the estimated CSM because the hydrodynamic noise is partially coherent and typically dominates the CSM. The main advantage of LO is that due to the initial guess, the number of true sources is known, which makes the post-processing integration of source objects obsolete, and the optimization process much faster (GO takes approximately 100x longer for the used SciPy implementations~\citep{2020SciPy-NMeth}). Since the initial source objects are initialized with the error $\sigma_x$, the results also show that the broadband energy is indeed smooth enough around the true positions for its gradient to lead towards the global minimum.

\subsection{Local optimization on synthetic multipoles}\label{sec:res_synth_dipole}
As described in the previous section it could be of interest to evaluate complex sources, by including parameterized properties in the propagation operator. One example is a multipole that contains poles of any order, such as a monopole, a dipole, and so on. We incorporate this by super-positioning the estimated, incoherent multipole CSMs in eq.~\ref{eq:cost_function} with
\begin{equation}
    c_\text{mod} = c_\text{monop.} + c_\text{dip.} + \dots,
\end{equation}
by using a shared location, and rotation angle for all frequencies and poles in a source object. A dipole has two additional variables, the rotation angle $\theta$, and $\varphi$. For $\theta=\varphi=0$, the dipole's main lobe is orientated in $x_3$ direction, $\theta$ turns the dipole around $x_2$, $\varphi$ around $x_3$.

\begin{table}[]
\centering
\begin{tabular}{c|ccc|c|ccc}
    var. & $x_1$ &$x_2$ &$x_3$ & pole& $Q(f)$ & $\theta$ & $\varphi$\\
    unit & \multicolumn{3}{c|}{$\si{\metre}$} & & $\si{\decibel\per\hertz}$ & \multicolumn{2}{c}{rad}\\
    \hline
    \multirow{2}{*}{$S_{I}$} & \multirow{2}{*}{0.5}& \multirow{2}{*}{0.5}& \multirow{2}{*}{0.0}& monop.& 100&- &-\\
    & & & & dip.& 60& $\pi/2$& $0$\\
    \multirow{2}{*}{$S_{II}$} & \multirow{2}{*}{0.0}& \multirow{2}{*}{0.5}& \multirow{2}{*}{0.0}& monop.& $-\infty$& -&-\\
    & & & & dip.& 40& $\pi/2$&$\pi/2$\\
\end{tabular}
\caption{Multipole parameters for case $6)$. The multipoles' amplitudes are constant over frequency.}
\label{tab:case6_param}
\end{table}

\begin{figure}[h!]
    \centering
    \includegraphics[width=3.48in]{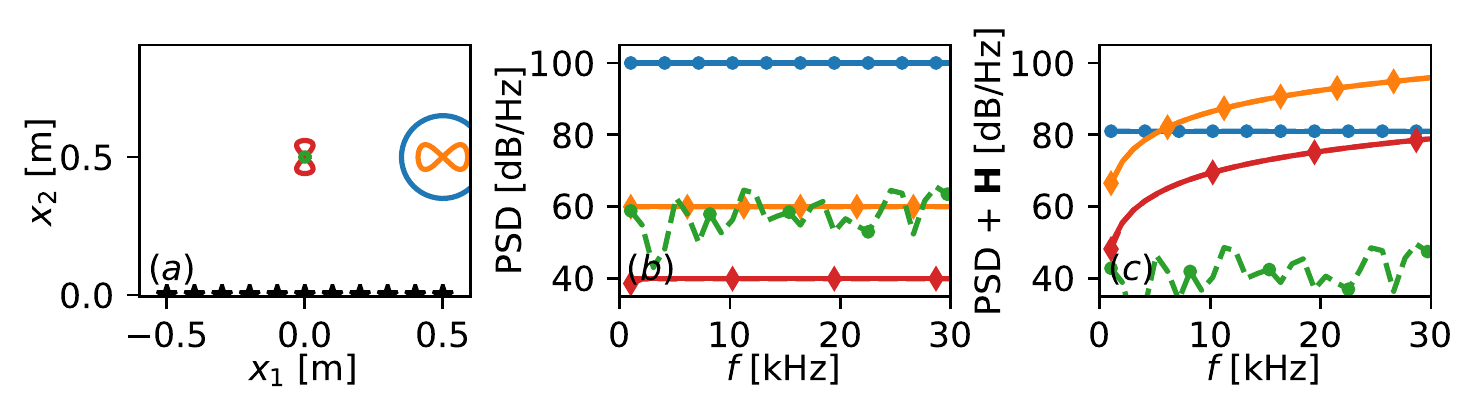}
    \caption{(Color online) Broadband LO result for case~$6)$ with $N_\text{true}=2$ multipole sources, consisting of incoherent monopoles (blue and green), and dipoles (orange and red). $(a)$ shows the true spatial location of the sources. The radius of the source shape is proportional to their strength. $(b)$ shows the broadband LO ($N_\text{est.}=2$) spectral estimation with ground truth (solid) and the estimation (dashed lines) for the monopoles ($\cdot$) and dipoles ($\lozenge$).  $(c)$ shows the perceived PSD at the array location due to the Green's functions eq.~\ref{eq:monopole_green} and eq.~\ref{eq:dipole_green} which indicates the pole's contribution to the observed CSM.}
    \label{fig:Figure17}
\end{figure}

As proof of concept, we perform LO for two synthetic multiples, consisting of monopoles and dipoles, with the parameters given in Table~\ref{tab:case6_param} with the array setup from Sec.~\ref{sec:GO_synthetic}. The setup is visualized in Figure~\ref{fig:Figure17} $(a)$. For LO, the initial positional start values are chosen according to the last section with $\sigma_x=0.025$m. The start amplitude for the optimizer is $\text{PSD}(f)=\SI{80}{\decibel\per\hertz}$. The start values for the dipole rotation angles $\varphi$ are random since there cannot be obtained reasonable start values from \replaced[R2C08]{naive}{forward} beamforming methods. $\theta=\pi/2$ is fixed, as it cannot be obtained with an 1D array.

Figure~\ref{fig:Figure17} shows the results for case $6)$. $(b)$ shows the true and estimated pole PSDs. The true and estimated PSDs coincide, except for $S_{II}$ where a monopole amplitude around $Q\approx\SI{60}{\decibel\per\hertz}$ is estimated, whereas the true amplitude is $Q= -\infty\si{\decibel\per\hertz}$. Note, that the SNR of this estimation appears distorted in $(b)$, since the monopole and dipole contribute differently to the total CSM, due to their different Green's functions~\hbox{\citep{Suzuki2011}}, see eq.~\ref{eq:monopole_green} and eq.~\ref{eq:dipole_green}. For reference, Figure~\ref{fig:Figure17} $(c)$ shows the perceived PSD at the array's position $x=[0,0,0]^T$, based on the Green's Matrix $\mathbf{H}$. The dipoles' rotation angles are estimated with errors below $\epsilon_{\varphi}<10^{-5}$ rad.

\section{Conclusion}\label{sec:conclusion}
This paper presented Global and Local Optimization for acoustic broadband sources, by introducing source objects with parameters such as the spatial location or source orientation, shared for all frequencies. The advantage over the single-frequency optimization process is the increased ratio of equations to unknown variables, the smoothing of local minima in the energy function, and the simple process of extracting the source positions and spectra from the results. 

While the approach outperformed methods such as CLEAN-SC, and standard Global Optimization on synthetic data, the results were similar on a real-world problem, containing three monopoles, with the proposed methods having a slight advantage in high-frequency regions. The main difference is the handling of noise and the violation of the monopole assumption due to the speaker. While CLEAN-SC results in strong noise scattered around the map, Global Optimization includes these violations in the estimated source spectra, which results in wrong PSD \added{estimations}. Thus, \replaced[R2C08]{naive}{forward} beamformers seem to be more forgiving towards violated assumptions, but Global Optimization provides the remaining energy as a measure to indicate the severeness of the violation.

One challenge for broadband Global Optimization is the correct determination of the number of source objects. It was shown, that an overestimation is a valid approach to finding all sources, or to accounting for properties not included in the Green's functions, such as reflections, or distributed sources. Then, multiple source objects approximate a single source. However, they can be identified based on their spatial distance and integrated much simpler than CLEAN-SC or standard Global Optimization results.

Since Global Optimization is computationally expensive and global optimizers are not guaranteed to find the global minimum, it was shown that by using reasonable starting locations for the source objects the problem can be reduced to Local Optimization since the combination of multiple frequencies reduces local minima in the energy function. A reason to perform Local or Global Optimization after obtaining reasonable initial values might be to identify source properties that are neglected by grid-based beamformers, such as multipoles, distributed sources, \added[R2C05]{coherent sources}, etc. The paper proposed a synthetic multipole problem which was reconstructed with low errors. 

\bibliography{main}{}

\begin{thebibliography}{10}
\def\enquote#1,{``#1,''}
\def\enxquote#1{``#1''}
\expandafter\ifx\csname url\endcsname\relax
  \def\url#1{\texttt{#1}}\fi
\expandafter\ifx\csname urlprefix\endcsname\relax\def\urlprefix{URL }\fi
\providecommand{\bibinfo}[2]{#2}
\def\plainquote#1{``#1''}
\providecommand{\noopsort}[1]{}
\providecommand{\switchargs}[2]{#2#1}
\providecommand{\dourl}[1]{\href{http://#1}{\nolinkurl{#1}}}
  \def\eatspace #1{#1}

\bibitem{MerinoMartinez2019a}
\bibinfo{author}{R.~Merino-Mart{\'i}nez}, \bibinfo{author}{P.~Sijtsma},
  \bibinfo{author}{M.~Snellen}, \bibinfo{author}{T.~Ahlefeldt},
  \bibinfo{author}{J.~Antoni}, \bibinfo{author}{C.~J. Bahr},
  \bibinfo{author}{D.~Blacodon}, \bibinfo{author}{D.~Ernst},
  \bibinfo{author}{A.~Finez}, \bibinfo{author}{S.~Funke},
  \bibinfo{author}{T.~F. Geyer}, \bibinfo{author}{S.~Haxter},
  \bibinfo{author}{G.~Herold}, \bibinfo{author}{X.~Huang},
  \bibinfo{author}{W.~M. Humphreys}, \bibinfo{author}{Q.~Lecl{\`e}re},
  \bibinfo{author}{A.~Malgoezar}, \bibinfo{author}{U.~Michel},
  \bibinfo{author}{T.~Padois}, \bibinfo{author}{A.~Pereira},
  \bibinfo{author}{C.~Picard}, \bibinfo{author}{E.~Sarradj},
  \bibinfo{author}{H.~Siller}, \bibinfo{author}{D.~G. Simons}, and
  \bibinfo{author}{C.~Spehr}, \enquote{\bibinfo{title}{A review of acoustic
  imaging methods using phased microphone arrays}},  \bibinfo{journal}{CEAS
  Aeronautical Journal} \textbf{10}(1), \bibinfo{pages}{197--230}
  (\bibinfo{year}{2019}) \dodoi{10.1007/s13272-019-00383-4}.

\bibitem{Hohage2020}
\bibinfo{author}{T.~Hohage}, \bibinfo{author}{H.-G. Raumer}, and
  \bibinfo{author}{C.~Spehr}, \enquote{\bibinfo{title}{Uniqueness of an inverse
  source problem in experimental aeroacoustics}},  \bibinfo{journal}{Inverse
  Problems} \textbf{36}(7), \bibinfo{pages}{075012} (\bibinfo{year}{2020})
  \dodoi{10.1088/1361-6420/ab8484}.

\bibitem{Ahlefeld2010}
\bibinfo{author}{T.~Ahlefeldt}, \bibinfo{author}{L.~Koop},
  \bibinfo{author}{A.~Lauterbach}, and \bibinfo{author}{C.~Spehr},
  \enquote{\bibinfo{title}{{A}dvances in microphone array measurements in a
  cryogenic wind tunnel}}, in \emph{\bibinfo{booktitle}{Berlin Beamforming
  Conference}} (\bibinfo{year}{2010}).

\bibitem{Bahr2017}
\bibinfo{author}{C.~J. Bahr}, \bibinfo{author}{W.~M. Humphreys},
  \bibinfo{author}{D.~Ernst}, \bibinfo{author}{T.~Ahlefeldt},
  \bibinfo{author}{C.~Spehr}, \bibinfo{author}{A.~Pereira},
  \bibinfo{author}{Q.~Leclère}, \bibinfo{author}{C.~Picard},
  \bibinfo{author}{R.~Porteous}, \bibinfo{author}{D.~Moreau},
  \bibinfo{author}{J.~R. Fischer}, and \bibinfo{author}{C.~J. Doolan},
  \enquote{\bibinfo{title}{A comparison of microphone phased array methods
  applied to the study of airframe noise in wind-tunnel testing}}, in
  \emph{\bibinfo{booktitle}{23rd AIAA/CEAS Aeroacoustics Conference}}
  (\bibinfo{year}{2017}), \dodoi{10.2514/6.2017-3718}.

\bibitem{Martinez2020}
\bibinfo{author}{R.~Merino-Martinez}, \bibinfo{author}{G.~Herold},
  \bibinfo{author}{M.~Snellen}, and \bibinfo{author}{R.~Dougherty},
  \enquote{\bibinfo{title}{Assessment and comparison of the performance of
  functional projection beamforming for aeroacoustic measurements}}, in
  \emph{\bibinfo{booktitle}{Berlin Beamforming Conference}}
  (\bibinfo{year}{2020}).

\bibitem{Lehmann2022}
\bibinfo{author}{M.~Lehmann}, \bibinfo{author}{D.~Ernst},
  \bibinfo{author}{M.~Schneider}, \bibinfo{author}{C.~Spehr}, and
  \bibinfo{author}{M.~Lummer}, \enquote{\bibinfo{title}{Beamforming for
  measurements under disturbed propagation conditions using numerically
  calculated green's functions}},  \bibinfo{journal}{Journal of Sound and
  Vibration} \textbf{520}, \bibinfo{pages}{116638} (\bibinfo{year}{2022})
  \dodoi{10.1016/j.jsv.2021.116638}.

\bibitem{Cox1987}
\bibinfo{author}{H.~Cox}, \bibinfo{author}{R.~Zeskind}, and
  \bibinfo{author}{M.~Owen}, \enquote{\bibinfo{title}{Robust adaptive
  beamforming}},  \bibinfo{journal}{{IEEE} Transactions on Acoustics, Speech,
  and Signal Processing} \textbf{35}(10), \bibinfo{pages}{1365--1376}
  (\bibinfo{year}{1987}) \dodoi{10.1109/tassp.1987.1165054}.

\bibitem{Suzuki2011}
\bibinfo{author}{T.~Suzuki}, \enquote{\bibinfo{title}{L1 generalized inverse
  beam-forming algorithm resolving coherent/incoherent, distributed and
  multipole sources}},  \bibinfo{journal}{Journal of Sound and Vibration}
  \textbf{330}(24), \bibinfo{pages}{5835--5851} (\bibinfo{year}{2011})
  \dodoi{10.1016/j.jsv.2011.05.021}.

\bibitem{Zavala2011}
\bibinfo{author}{P.~Zavala}, \bibinfo{author}{W.~D. Roeck},
  \bibinfo{author}{K.~Janssens}, \bibinfo{author}{J.~Arruda},
  \bibinfo{author}{P.~Sas}, and \bibinfo{author}{W.~Desmet},
  \enquote{\bibinfo{title}{Generalized inverse beamforming with optimized
  regularization strategy}},  \bibinfo{journal}{Mechanical Systems and Signal
  Processing} \textbf{25}(3), \bibinfo{pages}{928--939} (\bibinfo{year}{2011})
  \dodoi{10.1016/j.ymssp.2010.09.012}.

\bibitem{Sijtsma2007}
\bibinfo{author}{P.~Sijtsma}, \enquote{\bibinfo{title}{Clean based on spatial
  source coherence. international journal of aeroacoustics}},
  \bibinfo{journal}{International Journal of Aeroacoustics} \textbf{6},
  \bibinfo{pages}{357--374} (\bibinfo{year}{2007})
  \dodoi{10.1260/147547207783359459}.

\bibitem{Brooks2006}
\bibinfo{author}{T.~F. Brooks} and \bibinfo{author}{W.~M. Humphreys},
  \enquote{\bibinfo{title}{A deconvolution approach for the mapping of acoustic
  sources ({DAMAS}) determined from phased microphone arrays}},
  \bibinfo{journal}{Journal of Sound and Vibration} \textbf{294}(4-5),
  \bibinfo{pages}{856--879} (\bibinfo{year}{2006})
  \dodoi{10.1016/j.jsv.2005.12.046}.

\bibitem{Chardon2021b}
\bibinfo{author}{G.~Chardon}, \bibinfo{author}{J.~Picheral}, and
  \bibinfo{author}{F.~Ollivier}, \enquote{\bibinfo{title}{Theoretical analysis
  of the damas algorithm and efficient implementation of the covariance matrix
  fitting method for large-scale problems}},  \bibinfo{journal}{Journal of
  Sound and Vibration} \textbf{508}, \bibinfo{pages}{116208}
  (\bibinfo{year}{2021}) \dodoi{10.1016/j.jsv.2021.116208}.

\bibitem{Edelmann2011}
\bibinfo{author}{G.~F. Edelmann} and \bibinfo{author}{C.~F. Gaumond},
  \enquote{\bibinfo{title}{Beamforming using compressive sensing}},
  \bibinfo{journal}{The Journal of the Acoustical Society of America}
  \textbf{130}(4), \bibinfo{pages}{EL232--EL237} (\bibinfo{year}{2011})
  \dodoi{10.1121/1.3632046}.

\bibitem{Xenaki2014}
\bibinfo{author}{A.~Xenaki}, \bibinfo{author}{P.~Gerstoft}, and
  \bibinfo{author}{K.~Mosegaard}, \enquote{\bibinfo{title}{Compressive
  beamforming}},  \bibinfo{journal}{The Journal of the Acoustical Society of
  America} \textbf{136}(1), \bibinfo{pages}{260--271} (\bibinfo{year}{2014})
  \dodoi{10.1121/1.4883360}.

\bibitem{Chi2011}
\bibinfo{author}{Y.~Chi}, \bibinfo{author}{L.~L. Scharf},
  \bibinfo{author}{A.~Pezeshki}, and \bibinfo{author}{A.~R. Calderbank},
  \enquote{\bibinfo{title}{Sensitivity to basis mismatch in compressed
  sensing}},  \bibinfo{journal}{IEEE Trans. Signal Process.} \textbf{59},
  \bibinfo{pages}{2182} (\bibinfo{year}{2011}).

\bibitem{Duval2017}
\bibinfo{author}{V.~Duval} and \bibinfo{author}{G.~Peyr{\'{e}}},
  \enquote{\bibinfo{title}{Sparse spikes super-resolution on thin grids {II}:
  the continuous basis pursuit}},  \bibinfo{journal}{Inverse Problems}
  \textbf{33}(9), \bibinfo{pages}{095008} (\bibinfo{year}{2017})
  \dodoi{10.1088/1361-6420/aa7fce}.

\bibitem{Blacodon2004}
\bibinfo{author}{D.~Blacodon} and \bibinfo{author}{G.~Elias},
  \enquote{\bibinfo{title}{Level estimation of extended acoustic sources using
  a parametric method}},  \bibinfo{journal}{Journal of Aircraft}
  \textbf{41}(6), \bibinfo{pages}{1360--1369} (\bibinfo{year}{2004})
  \dodoi{10.2514/1.3053}.

\bibitem{Funke2012}
\bibinfo{author}{S.~Funke}, \bibinfo{author}{A.~Skorpel}, and
  \bibinfo{author}{U.~Michel}, \enquote{\bibinfo{title}{An extended formulation
  of the {SODIX} method with application to aeroengine broadband noise}}, in
  \emph{\bibinfo{booktitle}{18th {AIAA}/{CEAS} Aeroacoustics Conference (33rd
  {AIAA} Aeroacoustics Conference)}}, \bibinfo{publisher}{American Institute of
  Aeronautics and Astronautics} (\bibinfo{year}{2012}),
  \dodoi{10.2514/6.2012-2276}.

\bibitem{Oertwig2022}
\bibinfo{author}{S.~Oertwig}, \bibinfo{author}{H.~Siller}, and
  \bibinfo{author}{S.~Funke}, \enquote{\bibinfo{title}{Sodix for fully and
  partially coherent sound sources}}, in \emph{\bibinfo{booktitle}{Berlin
  Beamforming Conference}} (\bibinfo{year}{2022}).

\bibitem{Sarradj2022}
\bibinfo{author}{E.~Sarradj}, \enquote{\bibinfo{title}{Three-dimensional
  gridless source mapping using a signal subspace approach}}, in
  \emph{\bibinfo{booktitle}{Berlin Beamforming Conference}}
  (\bibinfo{year}{2022}).

\bibitem{Kujawski2022}
\bibinfo{author}{A.~Kujawski} and \bibinfo{author}{E.~Sarradj},
  \enquote{\bibinfo{title}{Fast grid-free strength mapping of multiple sound
  sources from microphone array data using a transformer architecture}},
  \bibinfo{journal}{The Journal of the Acoustical Society of America}
  \textbf{152}(5), \bibinfo{pages}{2543--2556} (\bibinfo{year}{2022})
  \dodoi{10.1121/10.0015005}.

\bibitem{Chardon2021a}
\bibinfo{author}{G.~Chardon} and \bibinfo{author}{U.~Boureau},
  \enquote{\bibinfo{title}{Gridless three-dimensional compressive beamforming
  with the sliding frank-wolfe algorithm}},  \bibinfo{journal}{The Journal of
  the Acoustical Society of America} \textbf{150}(4),
  \bibinfo{pages}{3139--3148} (\bibinfo{year}{2021})
  \dodoi{10.1121/10.0006790}.

\bibitem{Chardon2022a}
\bibinfo{author}{G.~Chardon}, \enquote{\bibinfo{title}{Gridless beamforming:
  Theoretical analysis of one source case, and sparsity based methods for
  multiple sources}}, in \emph{\bibinfo{booktitle}{Berlin Beamforming
  Conference}} (\bibinfo{year}{2022}).

\bibitem{Chardon2023}
\bibinfo{author}{G.~Chardon}, \enquote{\bibinfo{title}{Gridless covariance
  matrix fitting methods for three dimensional acoustical source
  localization}},  \bibinfo{journal}{Journal of Sound and Vibration}
  \textbf{551}, \bibinfo{pages}{117608} (\bibinfo{year}{2023})
  \dodoi{10.1016/j.jsv.2023.117608}.

\bibitem{Malgoezar2017}
\bibinfo{author}{A.~M.~N. Malgoezar}, \bibinfo{author}{M.~Snellen},
  \bibinfo{author}{R.~Merino-Martinez}, \bibinfo{author}{D.~G. Simons}, and
  \bibinfo{author}{P.~Sijtsma}, \enquote{\bibinfo{title}{On the use of global
  optimization methods for acoustic source mapping}},  \bibinfo{journal}{The
  Journal of the Acoustical Society of America} \textbf{141}(1),
  \bibinfo{pages}{453--465} (\bibinfo{year}{2017}) \dodoi{10.1121/1.4973915}.

\bibitem{Hoff2022}
\bibinfo{author}{B.~von~den Hoff}, \bibinfo{author}{R.~Merino-Mart{\'{\i}}nez},
  \bibinfo{author}{D.~G. Simons}, and \bibinfo{author}{M.~Snellen},
  \enquote{\bibinfo{title}{Using global optimization methods for
  three-dimensional localization and quantification of incoherent acoustic
  sources}},  \bibinfo{journal}{{JASA} Express Letters} \textbf{2}(5),
  \bibinfo{pages}{054802} (\bibinfo{year}{2022}) \dodoi{10.1121/10.0010456}.

\bibitem{Neumaier2004}
\bibinfo{author}{A.~Neumaier}, \enquote{\bibinfo{title}{Complete search in
  continuous global optimization and constraint satisfaction}},
  \bibinfo{journal}{Acta Numerica} \textbf{13}, \bibinfo{pages}{271–369}
  (\bibinfo{year}{2004}) \dodoi{10.1017/S0962492904000194}.

\bibitem{Dong2016}
\bibinfo{author}{B.~Dong}, \bibinfo{author}{J.~Antoni},
  \bibinfo{author}{A.~Pereira}, and \bibinfo{author}{W.~Kellermann},
  \enquote{\bibinfo{title}{Blind separation of incoherent and spatially
  disjoint sound sources}},  \bibinfo{journal}{Journal of Sound and Vibration}
  \textbf{383}, \bibinfo{pages}{414--445} (\bibinfo{year}{2016})
  \dodoi{10.1016/j.jsv.2016.07.018}.

\bibitem{HaxterSpehr2012}
\bibinfo{author}{S.~Haxter} and \bibinfo{author}{C.~Spehr},
  \enquote{\bibinfo{title}{Two-dimensional evaluation of turbulent boundary
  layer pressure fluctuations at cruise flight conditions}}, in
  \emph{\bibinfo{booktitle}{18th {AIAA}/{CEAS} Aeroacoustics Conference (33rd
  {AIAA} Aeroacoustics Conference)}}, \bibinfo{publisher}{American Institute of
  Aeronautics and Astronautics ({AIAA})} (\bibinfo{year}{2012}),
  \dodoi{10.2514/6.2012-2139}.

\bibitem{Haxter2014}
\bibinfo{author}{S.~Haxter} and \bibinfo{author}{C.~Spehr},
  \enquote{\bibinfo{title}{Infinite beamforming: Wavenumber decomposition of
  surface pressure fluctuations}}, in \emph{\bibinfo{booktitle}{Berlin
  Beamforming Conference}} (\bibinfo{year}{2014}).

\bibitem{Sijtsma2014}
\bibinfo{author}{P.~Sijtsma}, \bibinfo{author}{S.~Oerlemans},
  \bibinfo{author}{T.~G. Tibbe}, \bibinfo{author}{T.~Berkefeld}, and
  \bibinfo{author}{C.~Spehr}, \enquote{\bibinfo{title}{Spectral broadening by
  shear layers of open jet wind tunnels}}, in \emph{\bibinfo{booktitle}{20th
  {AIAA}/{CEAS} Aeroacoustics Conference}}, \bibinfo{publisher}{American
  Institute of Aeronautics and Astronautics ({AIAA})} (\bibinfo{year}{2014}),
  \dodoi{10.2514/6.2014-3178}.

\bibitem{ErnstSpehrBerkefeld2015}
\bibinfo{author}{D.~Ernst}, \bibinfo{author}{C.~Spehr}, and
  \bibinfo{author}{T.~Berkefeld}, \enquote{\bibinfo{title}{Decorrelation of
  acoustic wave propagation through the shear layer in open jet wind tunnel}},
  in \emph{\bibinfo{booktitle}{21st {AIAA}/{CEAS} Aeroacoustics Conference}},
  \bibinfo{publisher}{American Institute of Aeronautics and Astronautics
  ({AIAA})} (\bibinfo{year}{2015}), \dodoi{10.2514/6.2015-2976}.

\bibitem{Ernst2020}
\bibinfo{author}{D.~Ernst}, \enquote{\bibinfo{title}{Akustischer
  kohärenzverlust in offenen windkanälen aufgrund der turbulenten
  scherschicht}}, Ph.D. thesis, \bibinfo{school}{Technische Universität
  Berlin}, \bibinfo{year}{2020}, \dodoi{10.14279/DEPOSITONCE-9712}.

\bibitem{Yardibi2010}
\bibinfo{author}{T.~Yardibi}, \bibinfo{author}{J.~Li},
  \bibinfo{author}{P.~Stoica}, \bibinfo{author}{N.~S. Zawodny}, and
  \bibinfo{author}{L.~N. Cattafesta}, \enquote{\bibinfo{title}{A covariance
  fitting approach for correlated acoustic source mapping}},
  \bibinfo{journal}{The Journal of the Acoustical Society of America}
  \textbf{127}(5), \bibinfo{pages}{2920--2931} (\bibinfo{year}{2010})
  \dodoi{10.1121/1.3365260}.

\bibitem{MohanK2017}
\bibinfo{author}{D.~Mohan}, \bibinfo{author}{K.~N.},
  \bibinfo{author}{P.~Manovski}, \bibinfo{author}{R.~Geisler},
  \bibinfo{author}{J.~Agocs}, \bibinfo{author}{T.~Ahlefeldt},
  \bibinfo{author}{M.~Novara}, \bibinfo{author}{D.~Schanz},
  \bibinfo{author}{S.~Haxter}, \bibinfo{author}{D.~Ernst},
  \bibinfo{author}{C.~Spehr}, and \bibinfo{author}{A.~Schroeder},
  \enquote{\bibinfo{title}{Aeroacoustic analysis of a mach 0.9 round jet using
  synchronized microphone array and shake-the-box 3d lagrangian particle
  tracking measurements}}, in \emph{\bibinfo{booktitle}{AIAA AVIATION Forum}},
  \bibinfo{publisher}{American Institute of Aeronautics and Astronautics}
  (\bibinfo{year}{2017}), \dodoi{10.2514/6.2017-3862}.

\bibitem{Leclere2017}
\bibinfo{author}{Q.~Lecl{\`{e}}re}, \bibinfo{author}{A.~Pereira},
  \bibinfo{author}{C.~Bailly}, \bibinfo{author}{J.~Antoni}, and
  \bibinfo{author}{C.~Picard}, \enquote{\bibinfo{title}{A unified formalism for
  acoustic imaging based on microphone array measurements}},
  \bibinfo{journal}{International Journal of Aeroacoustics} \textbf{16}(4-5),
  \bibinfo{pages}{431--456} (\bibinfo{year}{2017})
  \dodoi{10.1177/1475472x17718883}.

\bibitem{Goudarzi_Bebec2022}
\bibinfo{author}{A.~Goudarzi}, \enquote{\bibinfo{title}{Frequency domain
  beamforming using neuronal networks}}, in \emph{\bibinfo{booktitle}{Berlin
  Beamforming Conference}} (\bibinfo{year}{2022}).

\bibitem{Chardon2022c}
\bibinfo{author}{G.~Chardon}, \enquote{\bibinfo{title}{Theoretical analysis of
  beamforming steering vector formulations for acoustic source localization}},
  \bibinfo{journal}{Journal of Sound and Vibration} \textbf{517},
  \bibinfo{pages}{116544} (\bibinfo{year}{2022})
  \dodoi{10.1016/j.jsv.2021.116544}.

\bibitem{AhlefeldtKoop2010}
\bibinfo{author}{T.~Ahlefeldt} and \bibinfo{author}{L.~Koop},
  \enquote{\bibinfo{title}{Microphone-array measurements in a cryogenic wind
  tunnel}},  \bibinfo{journal}{{AIAA} Journal} \textbf{48}(7),
  \bibinfo{pages}{1470--1479} (\bibinfo{year}{2010}) \dodoi{10.2514/1.j050083}.

\bibitem{Lauterbach_etal2010}
\bibinfo{author}{A.~Lauterbach}, \bibinfo{author}{K.~Ehrenfried},
  \bibinfo{author}{S.~Kröber}, \bibinfo{author}{T.~Ahlefeldt}, and
  \bibinfo{author}{S.~Loose}, \enquote{\bibinfo{title}{{M}icrophone array
  measurements on high-speed trains in wind tunnels}}, in
  \emph{\bibinfo{booktitle}{Berlin Beamforming Conference}}
  (\bibinfo{year}{2010}).

\bibitem{Ahlefeldt2021}
\bibinfo{author}{T.~Ahlefeldt}, \bibinfo{author}{S.~Haxter},
  \bibinfo{author}{C.~Spehr}, \bibinfo{author}{D.~Ernst}, and
  \bibinfo{author}{T.~Kleindienst}, \enquote{\bibinfo{title}{Road to
  acquisition: Preparing a {MEMS} microphone array for measurement of fuselage
  surface pressure fluctuations}},  \bibinfo{journal}{Micromachines}
  \textbf{12}(8), \bibinfo{pages}{961} (\bibinfo{year}{2021})
  \dodoi{10.3390/mi12080961}.

\bibitem{HaxterSpehr2016}
\bibinfo{author}{S.~Haxter} and \bibinfo{author}{C.~Spehr},
  \enquote{\bibinfo{title}{Comparison of model predictions for coherence length
  to in-flight measurements at cruise conditions}},  \bibinfo{journal}{Journal
  of Sound and Vibration}  (\bibinfo{year}{2016})
  \dodoi{10.1016/j.jsv.2016.10.038}.

\bibitem{Haxter201785}
\bibinfo{author}{S.~Haxter}, \bibinfo{author}{J.~Brouwer},
  \bibinfo{author}{J.~Sesterhenn}, and \bibinfo{author}{C.~Spehr},
  \enquote{\bibinfo{title}{Obtaining phase velocity of turbulent boundary layer
  pressure fluctuations at high subsonic mach number from wind tunnel data
  affected by strong background noise}},  \bibinfo{journal}{Journal of Sound
  and Vibration} \textbf{402}, \bibinfo{pages}{85 -- 103}
  (\bibinfo{year}{2017}) \dodoi{10.1016/j.jsv.2017.05.011}.

\bibitem{Ahlefeldt2013}
\bibinfo{author}{T.~Ahlefeldt}, \enquote{\bibinfo{title}{Aeroacoustic
  measurements of a scaled half-model at high reynolds numbers}},
  \bibinfo{journal}{{AIAA} Journal} \textbf{51}(12),
  \bibinfo{pages}{2783--2791} (\bibinfo{year}{2013}) \dodoi{10.2514/1.j052345}.

\bibitem{Ahlefeldt2016}
\bibinfo{author}{T.~Ahlefeldt}, \enquote{\bibinfo{title}{Microphone array
  measurement in european transonic wind tunnel at flight reynolds numbers}},
  \bibinfo{journal}{{AIAA} Journal} \bibinfo{pages}{1--13}
  (\bibinfo{year}{2016}) \dodoi{10.2514/1.j055262}.

\bibitem{Goudarzi2022}
\bibinfo{author}{A.~Goudarzi}, \enquote{\bibinfo{title}{Frequency domain
  beamforming using neuronal networks}}, in \emph{\bibinfo{booktitle}{Berlin
  Beamforming Conference}} (\bibinfo{year}{2022}).

\bibitem{Goudarzi2021}
\bibinfo{author}{A.~Goudarzi}, \bibinfo{author}{C.~Spehr}, and
  \bibinfo{author}{S.~Herbold}, \enquote{\bibinfo{title}{Automatic source
  localization and spectra generation from sparse beamforming maps}},
  \bibinfo{journal}{The Journal of the Acoustical Society of America}
  \textbf{150}(3), \bibinfo{pages}{1866--1882} (\bibinfo{year}{2021})
  \dodoi{10.1121/10.0005885}.

\bibitem{Brooks1999}
\bibinfo{author}{T.~Brooks} and \bibinfo{author}{J.~William~Humphreys},
  \enquote{\bibinfo{title}{Effect of directional array size on the measurement
  of airframe noise components}}, in \emph{\bibinfo{booktitle}{5th
  {AIAA}/{CEAS} Aeroacoustics Conference and Exhibit}},
  \bibinfo{publisher}{American Institute of Aeronautics and Astronautics}
  (\bibinfo{year}{1999}), \dodoi{10.2514/6.1999-1958}.

\bibitem{MerinoMartinez2019b}
\bibinfo{author}{R.~Merino-Mart{\'i}nez}, \bibinfo{author}{P.~Sijtsma},
  \bibinfo{author}{A.~Rubio~Carpio}, \bibinfo{author}{R.~Zamponi},
  \bibinfo{author}{S.~Luesutthiviboon}, \bibinfo{author}{A.~Malgoezar},
  \bibinfo{author}{M.~Snellen}, \bibinfo{author}{C.~Schram}, and
  \bibinfo{author}{D.~Simons}, \enquote{\bibinfo{title}{Integration methods for
  distributed sound sources}},  \bibinfo{journal}{International Journal of
  Aeroacoustics} \textbf{18}, \bibinfo{pages}{1475472X1985294}
  (\bibinfo{year}{2019}) \dodoi{10.1177/1475472X19852945}.

\bibitem{Vanderveen1996}
\bibinfo{author}{M.~Vanderveen}, \bibinfo{author}{B.~Ng},
  \bibinfo{author}{C.~Papadias}, and \bibinfo{author}{A.~Paulraj},
  \enquote{\bibinfo{title}{Joint angle and delay estimation ({JADE}) for
  signals in multipath environments}}, in \emph{\bibinfo{booktitle}{Conference
  Record of The Thirtieth Asilomar Conference on Signals, Systems and
  Computers}}, \bibinfo{publisher}{{IEEE} Comput. Soc. Press}
  (\bibinfo{year}{1996}), Vol.~\bibinfo{volume}{2}, pp.
  \bibinfo{pages}{1250--1254}, \dodoi{10.1109/acssc.1996.599145}.

\bibitem{Sarradj2012}
\bibinfo{author}{E.~Sarradj}, \enquote{\bibinfo{title}{Three-dimensional
  acoustic source mapping with different beamforming steering vector
  formulations}},  \bibinfo{journal}{Advances in Acoustics and Vibration}
  \textbf{2012}, \bibinfo{pages}{1--12} (\bibinfo{year}{2012})
  \dodoi{10.1155/2012/292695}.

\bibitem{Chardon2022b}
\bibinfo{author}{G.~Chardon}, \enquote{\bibinfo{title}{Theoretical analysis of
  beamforming steering vector formulations for acoustic source localization}},
  \bibinfo{journal}{Journal of Sound and Vibration} \textbf{517},
  \bibinfo{pages}{116544} (\bibinfo{year}{2022})
  \dodoi{10.1016/j.jsv.2021.116544}.

\bibitem{Simons2017}
\bibinfo{author}{D.~G. Simons}, \bibinfo{author}{M.~Snellen},
  \bibinfo{author}{R.~M. Mart{\'i}nez}, and \bibinfo{author}{A.~Malgoezar},
  \enquote{\bibinfo{title}{Noise breakdown of landing aircraft using a
  microphone array and an airframe noise model}}, in
  \emph{\bibinfo{booktitle}{Internoise}} (\bibinfo{year}{2017}).

\bibitem{McInnes2017}
\bibinfo{author}{L.~McInnes}, \bibinfo{author}{J.~Healy}, and
  \bibinfo{author}{S.~Astels}, \enquote{\bibinfo{title}{hdbscan: Hierarchical
  density based clustering}},  \bibinfo{journal}{The Journal of Open Source
  Software} \textbf{2}(11) (\bibinfo{year}{2017}) \dodoi{10.21105/joss.00205}.

\bibitem{2020SciPy-NMeth}
\bibinfo{author}{P.~Virtanen}, \bibinfo{author}{R.~Gommers},
  \bibinfo{author}{T.~E. Oliphant}, \bibinfo{author}{M.~Haberland},
  \bibinfo{author}{T.~Reddy}, \bibinfo{author}{D.~Cournapeau},
  \bibinfo{author}{E.~Burovski}, \bibinfo{author}{P.~Peterson},
  \bibinfo{author}{W.~Weckesser}, \bibinfo{author}{J.~Bright},
  \bibinfo{author}{S.~J. {van der Walt}}, \bibinfo{author}{M.~Brett},
  \bibinfo{author}{J.~Wilson}, \bibinfo{author}{K.~J. Millman},
  \bibinfo{author}{N.~Mayorov}, \bibinfo{author}{A.~R.~J. Nelson},
  \bibinfo{author}{E.~Jones}, \bibinfo{author}{R.~Kern},
  \bibinfo{author}{E.~Larson}, \bibinfo{author}{C.~J. Carey},
  \bibinfo{author}{{\.I}.~Polat}, \bibinfo{author}{Y.~Feng},
  \bibinfo{author}{E.~W. Moore}, \bibinfo{author}{J.~{VanderPlas}},
  \bibinfo{author}{D.~Laxalde}, \bibinfo{author}{J.~Perktold},
  \bibinfo{author}{R.~Cimrman}, \bibinfo{author}{I.~Henriksen},
  \bibinfo{author}{E.~A. Quintero}, \bibinfo{author}{C.~R. Harris},
  \bibinfo{author}{A.~M. Archibald}, \bibinfo{author}{A.~H. Ribeiro},
  \bibinfo{author}{F.~Pedregosa}, \bibinfo{author}{P.~{van Mulbregt}}, and
  \bibinfo{author}{{SciPy 1.0 Contributors}}, \enquote{\bibinfo{title}{{{SciPy}
  1.0: Fundamental Algorithms for Scientific Computing in Python}}},
  \bibinfo{journal}{Nature Methods} \textbf{17}, \bibinfo{pages}{261--272}
  (\bibinfo{year}{2020}) \dodoi{10.1038/s41592-019-0686-2}.

\bibitem{Amiet1976}
\bibinfo{author}{R.~K. Amiet}, \enquote{\bibinfo{title}{Correction of open-jet
  wind-tunnel measurements for shear layer refraction}}, in
  \emph{\bibinfo{booktitle}{Aeroacoustics: Acoustic Wave Propagation; Aircraft
  Noise Prediction; Aeroacoustic Instrumentation}}
  (\bibinfo{publisher}{American Institute of Aeronautics and Astronautics},
  \bibinfo{year}{1976}), pp. \bibinfo{pages}{259--280},
  \dodoi{10.2514/5.9781600865206.0259.0280}.

\end{thebibliography}
\end{document}